\documentclass[sigconf]{acmart}
\AtBeginDocument{%
  }

\copyrightyear{2025}
\acmYear{2025}
\setcopyright{cc}
\setcctype{by}
\acmConference[ASIA CCS '25]{ACM Asia Conference on Computer and Communications Security}{August 25--29, 2025}{Hanoi, Vietnam}
\acmBooktitle{ACM Asia Conference on Computer and Communications Security (ASIA CCS '25), August 25--29, 2025, Hanoi, Vietnam}\acmDOI{10.1145/3708821.3733867}
\acmISBN{979-8-4007-1410-8/2025/08}

\usepackage[T1]{fontenc}
\usepackage{url}
\usepackage{commath}
\usepackage{booktabs}  
\usepackage{amsfonts}     
\usepackage{nicefrac}     
\usepackage{microtype} 
\usepackage{multirow}
\usepackage{graphicx}
\usepackage{xcolor}
\usepackage{caption}
\usepackage{subcaption}
\usepackage{makecell}
\usepackage{array}
\usepackage{xspace}

\def\name{Minerva\xspace} 

\usepackage{eso-pic}

\AddToShipoutPictureBG{
  \AtPageUpperLeft{
    \raisebox{-2\baselineskip}{\makebox[\paperwidth]{\begin{minipage}{21cm}\centering
      Paper accepted at the 2025 ACM Asia Conference on Computer and Communications Security (ASIA CCS ’25),
    \end{minipage}}}
  }
}

\begin{document}
\title{\textsc{\name}: A File-Based Ransomware Detector}

\author{Dorjan Hitaj}
\affiliation{%
 \institution{Sapienza University of Rome}
  \country{Rome, Italy}
}
\orcid{0000-0001-5686-3831}
\email{hitaj.d@di.uniroma1.it}

\author{Giulio Pagnotta}
\affiliation{%
 \institution{Sapienza University of Rome}
  \country{Rome, Italy}
}
\orcid{0000-0002-4626-6045}
\email{pagnotta@di.uniroma1.it}

\author{Fabio De Gaspari}
\affiliation{%
 \institution{Sapienza University of Rome}
 \country{Rome, Italy}
}
\orcid{0000-0001-9718-1044}
\email{degaspari@di.uniroma1.it}

\author{Lorenzo De Carli}
\affiliation{%
 \institution{University of Calgary}
  \country{Calgary, Canada}
}
\orcid{0000-0003-0432-3686}
\email{lorenzo.decarli@ucalgary.ca}

\author{Luigi V. Mancini}
\affiliation{%
 \institution{Sapienza University of Rome}
  \country{Rome, Italy}
}
\orcid{0000-0003-4859-2191}
\email{mancini@di.uniroma1.it}

\renewcommand{\shortauthors}{Hitaj et al.}

\begin{abstract}
Ransomware attacks have caused billions of dollars in damages in recent years, and are expected to cause billions more in the future. Consequently, significant effort has been devoted to ransomware detection and mitigation. Behavioral-based ransomware detection approaches have garnered considerable attention recently.
These behavioral detectors typically rely on process-based behavioral profiles to identify malicious behaviors. However, with an increasing body of literature highlighting the vulnerability of such approaches to evasion attacks, a comprehensive solution to the ransomware problem remains elusive.

This paper presents \name, a novel robust approach to ransomware detection. \name is engineered to be robust by design against evasion attacks, with architectural and feature selection choices informed by their resilience to adversarial manipulation. We conduct a comprehensive analysis of \name across a diverse spectrum of ransomware types, encompassing unseen ransomware as well as variants designed specifically to evade \name. Our evaluation showcases the ability of \name to accurately identify ransomware, generalize to unseen threats, and withstand evasion attacks. Furthermore, over $99\%$ of detected ransomware are identified within $0.52sec$ of activity, enabling the adoption of data loss prevention techniques with near-zero overhead.
\end{abstract}

\keywords{ransomware detection, behavioral classification, machine learning}

\ccsdesc[500]{Security and privacy~Malware and its mitigation}
\ccsdesc[500]{Computing methodologies~Machine learning}

\maketitle

\section{Introduction}\label{sec:introduction}
Despite intense research~\cite{10.1145/3453153}, ransomware attacks continue to pose significant challenges. In ransomware attacks, a malicious actor infiltrates the systems of a victim organization (such as an enterprise~\cite{derek_kortepeter_shipping_2018} or a city government~\cite{blinder_cyberattack_2018}) and installs specialized malware---ransomware---that encrypts sensitive organization data. Subsequently, a ransom demand is made in exchange for decryption keys. Ransomware attacks can result in substantial financial losses for victims and disrupt critical systems, as evidenced by the recent Colonial Pipeline incident~\cite{cybersecurity_ventures,cybersecurity_web_page}.

The detection of ransomware on servers and user machines is a crucial last line of defense. Several detection algorithms can be employed to identify ransomware activity, terminate unwanted processes, and restore encrypted files to their original content~\cite{sheen2018ransomware,10430405,baek2018ssd}. Many existing approaches~\cite{continella_shieldfs:_2016,mehnaz_rwguard,kirda_redemption,10017139} leverage process-level features to differentiate ransomware processes performing file encryption from other benign processes executing on the machine. The most effective process-based detectors typically utilize behavioral detection: they construct behavioral profiles of running processes based on their system activity, and compare these profiles against a learned behavioral model of benign and ransomware processes. This approach is reasonable: ransomware operations involve intense disk I/O and encryption, typically resulting in distinct behavioral profiles compared to benign applications. Generally, these behavioral-based detectors rely on Machine Learning (ML) techniques to construct and compare behavioral profiles. ML techniques have proven to be valuable tools across various cybersecurity domains \cite{8907413,malphase,10.1145/3607199.3607207}, yielding substantial improvements over prior art~\cite{10.1145/3607199.3607206,9257172,pagnotta2022passflow}. However, recent research~\cite{de_gaspari_evading_2022,zhou2023limits} has revealed a series of evasive techniques that can readily bypass both academic and commercial process-based behavioral detectors. These evasive ransomware attacks entail distributing the ransomware workload across several cooperating processes, each emulating, feature-wise, the behavior of a benign process. Effectively, the behavioral profile of each ransomware process is indistinguishable from that of benign processes, but their coordinated action still results in the expected ransomware behavior. This evasive ransomware approach exploits inherent weaknesses of process-level features and, as such, cannot be identified at the process level. These attacks fall under a broader category of adversarial attacks against ML systems~\cite{rosenberg2021adversarial,10.1145/3607199.3607200,hitaj2022maleficnet}, which highlight the risks associated with such approaches.

\begin{figure}[t]
  \centering
  \includegraphics[width=\linewidth]{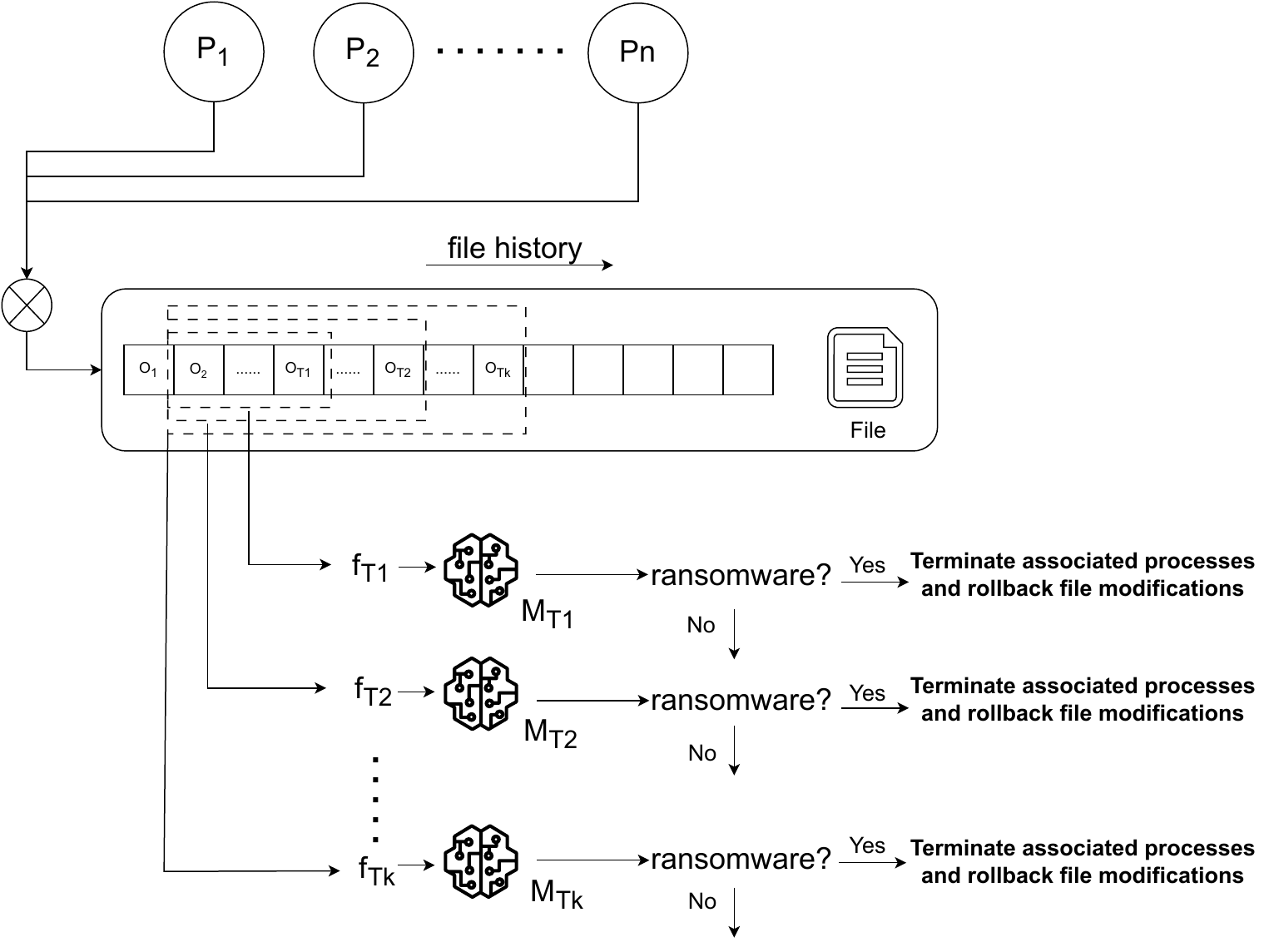}
  \caption{Overview of \name's ransomware detection module. For each individual file, \name maintains a record of operations performed by various processes \textbf{P} across different temporal resolutions (\textit{windows}). At the conclusion of each window, file operations are aggregated and submitted to the relevant ML classifier trained for that temporal resolution.}
  \label{fig:minerva_detection_module}
\end{figure}

This paper introduces \textbf{\name}, a novel defense against ransomware. Unlike prior methods that primarily rely on process-level behavioral analysis, \name classifies system activity by constructing file-level behavioral profiles for each file, based on the operations they undergo within a defined time window. Ransomware processes are identified based on their interaction with a file during a time window in which malicious activity is detected. We also engineer \name to be robust by design against complex evasion attacks by basing its architecture on a \textit{contrastive design}, where altering one aspect of the behavioral profile triggers detectable changes to other aspects. \name's architecture is based on two key insights. 
(1) The first insight is that while evasive ransomware attacks may mimic benign applications at the level of individual ransomware processes, the ultimate objective of ransomware is to encrypt user files. Therefore, regardless of how tasks are distributed among these individual processes, the ransomware's activity becomes evident from a file-based perspective.
(2) The second insight recognizes that different aspects of file-based behavioral profiles are interconnected and, due to constraints in the ransomware goals, changes to one aspect directly influence others. For instance, ransomware can manipulate the file-based behavioral profile of a file by altering the average entropy of file operations. However, to fully encrypt the file, it must eventually overwrite it with high-entropy data. To satisfy this constraint, ransomware must increase the number of operations on the file, introducing low-entropy operations to decrease average entropy, thereby modifying another aspect of the behavioral profile. Figure~\ref{fig:minerva_detection_module} provides a high-level overview of \name's detection module.

\noindent \textbf{Summarizing, this paper makes the following contributions:}

\begin{enumerate}
\item We propose \name, a robust ransomware detector based on the novel concepts of file-based behavioral profiles and contrastive design. \name overcomes critical limitations of existing behavioral detection techniques.

\item We perform a comprehensive analysis of the features generated by ransomware activity, identifying a robust set of features intrinsically related to core ransomware operations.

\item We assess \name's performance against traditional, evasive multiprocess, and unseen ransomware. Our findings demonstrate that \name effectively detects ransomware activity on average within $0.52$ seconds of the onset of malicious activity.

\item We analyze the resilience of \name to \emph{adaptive, unseen ransomware}: ransomware specifically engineered to evade \name's detection. We introduce three distinct families of adaptive ransomware and validate \name's robustness against each, emphasizing its capacity for generalization and underscoring the significance of our contrastive design.

\item We study the explainability of \name and provide key insights on its effectiveness.

\end{enumerate}

\section{Related Work}\label{sec:relatedwork}
Ransomware detection approaches rely on the insight that ransomware processes behave very differently from benign processes. Such approaches commonly consist of three main components. A monitoring component monitors the dynamic behavior of running processes, which is used to compute a set of \textit{features} based on program execution. Feature computation typically considers a sequence of recent process system calls whose characteristics are aggregated over a window of operations. Typical features used for this purpose are frequency and characteristics of file read/write operations, changes to file types (such as file type and file content entropy), and others. All these features are computed per process and then fed to a machine learning classifier, which uses them to label processes as benign or ransomware. Several techniques proposed in recent years follow the approach above: UNVEIL~\cite{kirda_unveil} and Redemption~\cite{kirda_redemption} generate a suspicious activity score based on characteristics of file write operations, entropy changes, and others. CryptoDrop~\cite{scaife_cryptolock_2016} generates a reputation score computed similarly. ShieldFS~\cite{continella_shieldfs:_2016} uses relative frequencies of various disk operations, such as read, write, folder listings, and others, as features fed to a random forest process classifier. A similar approach is also used by RWGuard~\cite{mehnaz_rwguard}. While detection techniques used in commercial detectors are not disclosed, De Gaspari et al.~\cite{naked_sun_acns} show that Malwarebytes's~\cite{malwarebytes_2019} ransomware detector is vulnerable to evasion attacks designed against process-based detectors and thus likely to leverage similar techniques.

Recently, the same authors proposed a novel family of evasion attacks against process-based ransomware detectors~\cite{de_gaspari_evading_2022}. These attacks fall under a broader category of adversarial attacks against ML systems~\cite{rosenberg2021adversarial,de2024have,hitaj2024you}, highlighting the risks associated with such approaches. The authors suggested three attacks: \textit{process splitting}, \textit{functional splitting}, and \textit{mimicry attack}. All these attacks aim to keep behavioral features' values below the detection threshold through various mechanisms, but the core principle is to split the ransomware operations over multiple independent processes. The \textit{process splitting} attack evenly distributes the malicious activity across several processes. This approach tends to be impractical in practice, requiring the creation of many processes. The \textit{functional splitting} attack spawns fewer processes, each performing a subset of ransomware operations. While this attack is practical, it can be mitigated with adversarial training~\cite{de_gaspari_evading_2022}. Finally, the \textit{mimicry attack} works by splitting ransomware into multiple processes, each mimicking the behavioral profile of benign applications from the point of view of disk operations. Mimicry attacks are fundamentally hard to detect as they are designed to resemble, feature-wise, the behavior of a benign process. \c{G}enc et al.~\cite{genc_next_2018} provides a high-level discussion of similar evasive approaches. MalWASH~\cite{malwash}, D-TIME~\cite{dtime}, and ROPE~\cite{delia_rope_2021} are generic malware obfuscators that also leverage the distribution of malware workload across processes, similarly to~\cite{de_gaspari_evading_2022}. Rather than distributing malware workload, other approaches attempt to perturb the sequence of system calls generated by ransomware~\cite{rosenberg_query-efficient_2018,hu_black-box_2017}.

Other approaches focus on using high entropy as a proxy to detect ransomware encryption activity. These approaches use various techniques, including $\chi^2$-test~\cite{lipmaa_data_2017} and the Kullback-Liebler divergence~\cite{foresti_efficient_2016}. However, when used in isolation, entropy is a weak distinguisher of ransomware activity and can be easily evaded~\cite{de_gaspari_reliable_2022}.

Another line of work utilizes decoy files~\cite{genc_deception_2019,moore_detecting_2016,moussaileb_ransomwares_2018}. A decoy file is a dummy file, typically hidden from the user and created for ransomware detection. As no application is expected to use a decoy file, accesses to it are a strong signal of ransomware activity.  
Recently, Ganfure et al.~\cite{10026355} presented RTrap, an advanced version of traditional decoy files. Rtrap is a machine learning-based approach to decoy files to provide early detection against cryptographic ransomware. In particular, RTrap selects a set of legitimate user files from the system, duplicates and deploys them throughout the system. 
The user files to be used as decoys are selected in such a way that they are representative of the user content on the machine. The insight behind the approach is that modern ransomware does not simply encrypt all files indiscriminately, but rather prioritizes certain types of file contents over others in order to be more effective. Panzade~\textit{et~al}~\cite{panzade2024vig} showed that RTrap can be evaded by not trying to encrypt the user files in a particular directory at the very beginning but rather, stealthily, ﬁnd out which files can potentially be a decoy and then build an encryption plan that skips them. 

In contrast to prior work on ransomware detection, \name was designed with the goal of being resilient to evasion attacks. This goal led to a detailed investigation of ransomware activity, from which we decided to monitor and model the behavior of the system activity from a file point-of-view. On top of that, the features selected for monitoring were chosen following a contrastive design paradigm, meaning that alternating one or more features would inevitably alter the rest of the remaining features, thus leaving little to no room for evasive ransomware to try to evade detection. The following section thoroughly introduces our approach and evaluates its ransomware detection capabilities. To the best of our knowledge, we are the first to fully evaluate a ransomware detection approach from an adaptive adversary's point-of-view, a feat not done in prior ransomware detection approaches.

\section{Threat Model}\label{sec:threatmodel}
We position ourselves in a threat model similar to those considered by prior work in the domain~\cite{continella_shieldfs:_2016,mehnaz_rwguard,kirda_redemption,10017139}: the adversary is a ransomware that has infected a target machine with the goal of encrypting users files and demanding a ransom to recover the data. We consider different types of ransomware, categorized into three groups according to their capabilities and behaviors:
\begin{itemize}
    \item \textbf{Traditional Ransomware,} corresponding to the typical ransomware families found in the wild.  
    \item \textbf{Evasive Multiprocess Ransomware,} corresponding to complex ransomware that mimics the behavior of benign processes to avoid detection~\cite{naked_sun_acns,zhou2023limits}.
    \item \textbf{Adaptive Ransomware,} corresponding to a sophisticated ransomware engineered specifically to evade \name's file-based detection through manipulation of features utilized by \name for detection.
\end{itemize}

\noindent

\textbf{Trusted Components.} 
We assume \name's Disk Activity Monitor (DAM) and File-Based Behavioral Detector (FBD) components to be trusted, meaning that ransomware processes cannot tamper with their functionality. The DAM module is a driver running in kernel space, which is protected by the operating system. The FBD is a user-space component that communicates with the DAM through a protected, read-only communication port. The communication's security context is set up so that only the system group associated with the FBD can access it, and the FBD runs with elevated (administrator) privileges to prevent tampering by other processes. We do not consider kernel-space ransomware, as malware with kernel-level access can effectively bypass any system-level countermeasure. These assumptions are in line with prior work in the field~\cite{continella_shieldfs:_2016,mehnaz_rwguard,kirda_redemption,10017139}. 

\section{\name: A File-Based Ransomware Detector}\label{sec:ourapproach}
This section presents \name, a novel ransomware detector built upon the concepts of file-based behavioral profiling and contrastive design. Minerva effectively guards against traditional ransomware, evasive multiprocess ransomware, and adaptive ransomware engineered specifically to evade \name's detection. \name's architecture is based on two key insights. 
(1) The first insight is that, although individual ransomware processes may modulate their behavior to imitate benign applications~\cite{de_gaspari_evading_2022,zhou2023limits}, they must ultimately encrypt large portions of users files to achieve their objective. Consequently, it is possible to monitor the behavioral profiles of the files themselves and detect any deviation from expected behaviors. From a file-based standpoint, the ransomware activity becomes apparent regardless of the number of processes performing encryption, or how tasks are distributed among individual ransomware processes. \name leverages this insight by creating file-based behavioral profiles for each opened user file in the system, based on the I/O operations they receive in a given time window. Ransomware detection is performed per file and per time window based on the constructed profiles.
(2) The second insight is that different aspects of file-based behavioral profiles are interconnected. Therefore, any attempt to alter some aspects of the behavioral profile to evade detection inevitably leads to detectable changes in others. \name capitalizes on this insight through its contrastive design: \name's features and architecture are selected so that any manipulation by ransomware to evade detection inevitably causes changes to other features that still expose the malicious activity, or that are captured by the tiered architecture of \name (see Section~\ref{sec:architecture}). We provide an in-depth discussion on \name's contrastive design in Section~\ref{sec:feature_analysis}.

\subsection{\name Architecture}
\label{sec:architecture}
From a high-level view, \name is comprised of two separate modules: (1) \emph{Disk Activity Monitor} (DAM), and (2) \emph{File-Based Behavioral Detector} (FBD). In what follows, we explain how these two modules work.

\begin{figure*}[t]
    \centering        
	    \begin{subfigure}{.24\textwidth}
            \centering
            \includegraphics[width=\columnwidth]{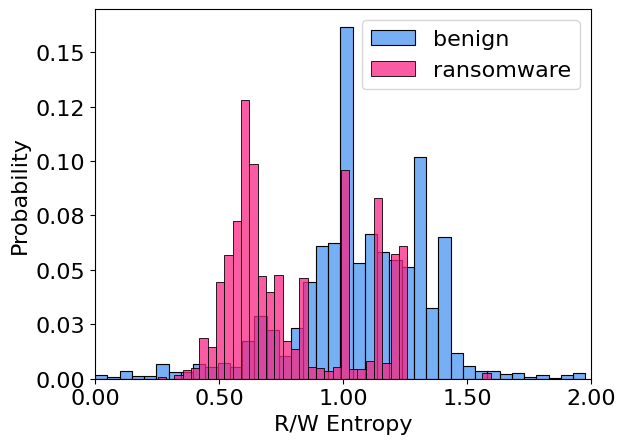}
             \caption{R/W ent. ratio.}
             \label{fig:hist_rw_ent}
        \end{subfigure}
        \hfill
	    \begin{subfigure}{.24\textwidth}
            \centering
            \includegraphics[width=\columnwidth]{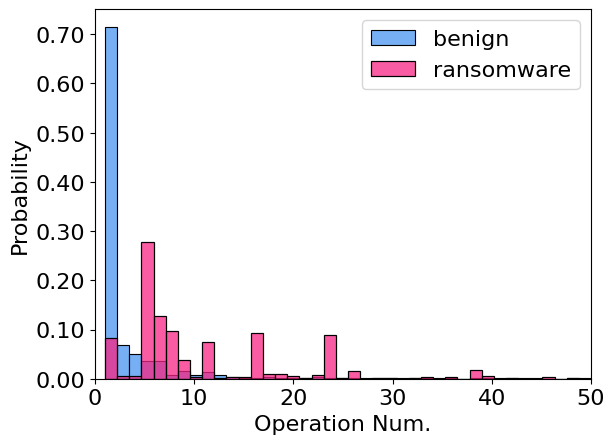}
             \caption{Num. of operations.}
             \label{fig:hist_op_num}
        \end{subfigure}
        \hfill
	    \begin{subfigure}{.24\textwidth}
            \centering
            \includegraphics[width=\columnwidth]{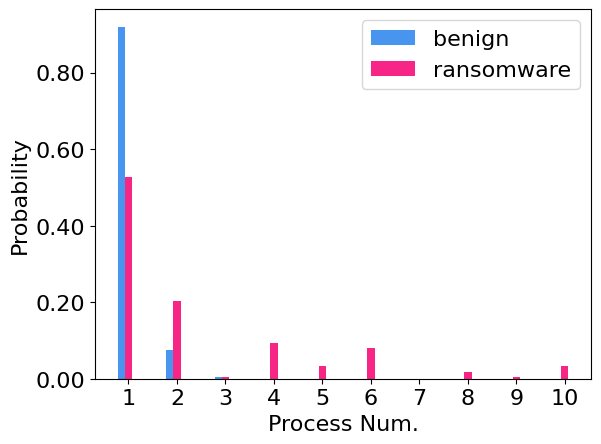}
             \caption{Num. of processes.}
             \label{fig:bar_proc_num}
        \end{subfigure}
        \hfill
	    \begin{subfigure}{.24\textwidth}
            \centering
            \includegraphics[width=\columnwidth]{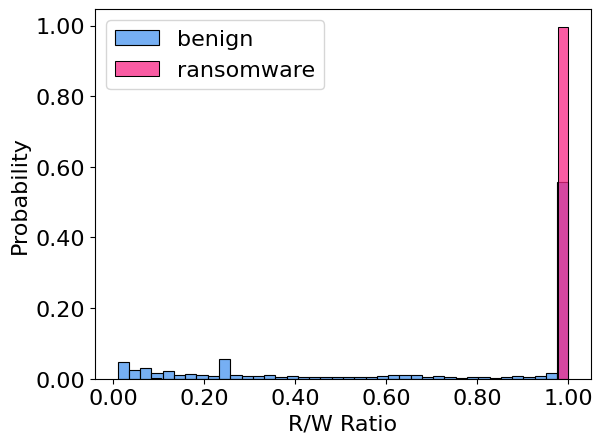}
             \caption{R/W ratio.}
             \label{fig:hist_rw_ratio}
        \end{subfigure}
	\begin{subfigure}{.24\textwidth}
            \centering
            \includegraphics[width=\columnwidth]{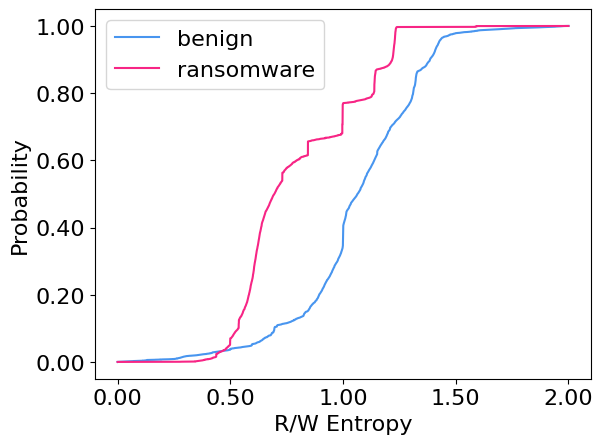}
             \caption{R/W ent. ratio CDF.}
             \label{fig:cdf_rw_ent}
        \end{subfigure}
        \hfill
	    \begin{subfigure}{.24\textwidth}
            \centering
            \includegraphics[width=\columnwidth]{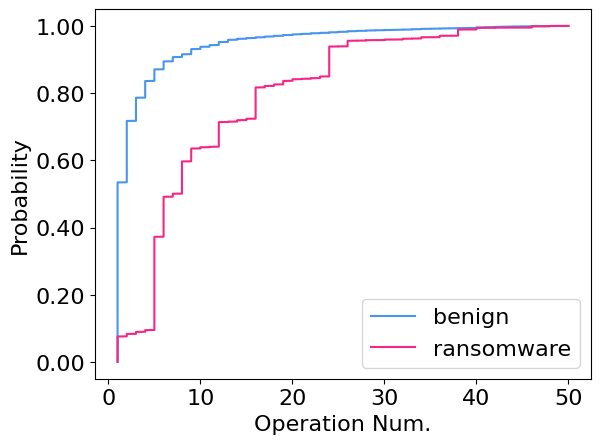}
             \caption{Num. of ops. CDF.}
             \label{fig:cdf_op_num}
        \end{subfigure}
        \hfill
	    \begin{subfigure}{.24\textwidth}
            \centering
            \includegraphics[width=\columnwidth]{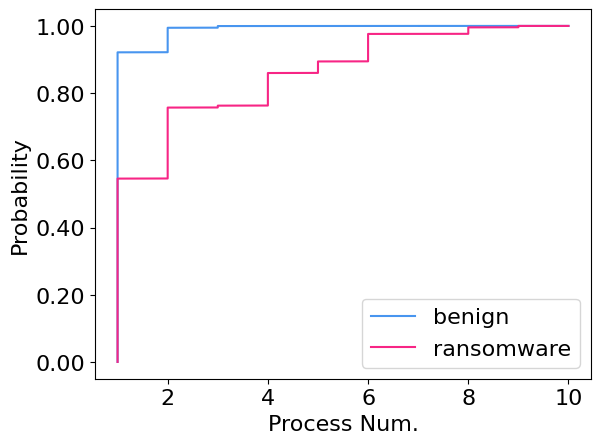}
             \caption{Num of Proc. CDF.}
             \label{fig:cdf_proc_num}
        \end{subfigure}
        \hfill
	    \begin{subfigure}{.24\textwidth}
            \centering
            \includegraphics[width=\columnwidth]{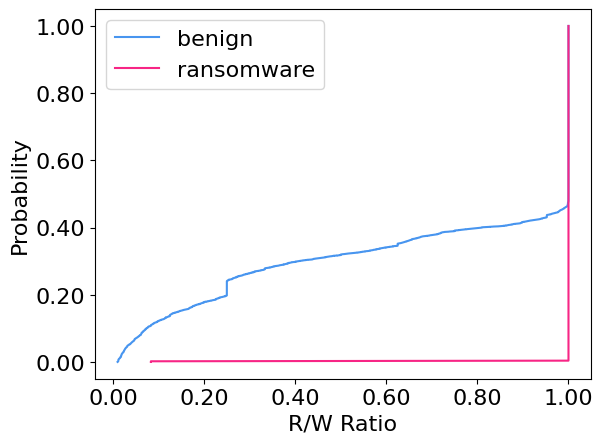}
             \caption{R/W ratio CDF.}
             \label{fig:cdf_rw_ratio}
        \end{subfigure}
    \caption{Comparison of distribution and Cumulative Distribution Function (CDF) of different Minerva features for benign and ransomware processes. Computed on 1 second window.} 
    \label{fig:feature_distribution_plots}
\end{figure*}

\subsubsection{Disk Activity Monitor.} \label{sec:disk_monitor}
The DAM module is a shim that provides system-wide, real-time information on all file system operations performed. We implemented the module as a driver that uses the I/O Request Packet (IRP) data structure provided by Windows systems to transparently log filesystem operations system-wide. Each filesystem-related system call performed by a process can result in many lower-level operations. The DAM captures and filters all these low-level operations before processing them into feature vectors, which are then forwarded to the FBD module of \name. 
To minimize overhead, we logically divide files into critical and non-critical: critical files encompass all user files that are high-priority targets for ransomware, while non-critical files comprise system and configuration files that can easily be recovered after an attack. \name automatically identifies as critical files all files located in non-system directories such as Documents, Downloads, and Desktop on Windows systems. System, program, temporary, and configuration files are excluded from monitoring. This includes all files in system directories (e.g., Windows) and in directories used by programs to store configuration and logging data (e.g., Program Files, ProgramData in Windows systems). \name's critical files criteria can be further customized by system administrators to include or exclude additional directories and files. The DAM excludes any operation performed on non-critical files and computes per-file feature vectors for all other opened files.

\subsubsection{File-Based Behavioral Detector.}\label{sec:detector_module} 
Figure~\ref{fig:minerva_detection_module} illustrates the architecture of the FBD module of \name. This module comprises an ensemble of machine-learning classifiers trained to identify ransomware activity by analyzing the behavioral profile of files. For each critical file in the system, \name computes several features based on the operations received by the file within a predefined time window. These features define that file's behavioral profile for that time window, which the Detector uses to determine whether the file has undergone malicious activity within the window. To withstand variations in ransomware behavior, such as adaptive attacks aimed at slowing ransomware activity to span multiple windows, \name employs a multi-tier architecture. Each tier implements a distinct fully-connected Deep Neural Network classifier trained on feature vectors computed over progressively longer time windows: low-tier classifiers utilize feature vectors computed over short time windows (e.g., $0.25$ sec), while high-tier classifiers utilize feature vectors calculated over longer windows (e.g., $4$ sec). At the end of each time window, \name queries the classifier of the corresponding tier to detect malicious behavioral profiles. In order to confirm malicious activity on a file, \name requires that at least one of its classifiers reports $K$ consecutive windows with detected malicious activity. If a classifier reports malicious activity for $K$ consecutive windows, all processes that were active on the specific file during those windows are flagged as ransomware processes. This procedure is carried out for each active critical file in the system that is receiving read or write operations. 

The architecture of \name's file-based detector is designed to accomplish two primary objectives. The first is to provide resilience to ransomware behavioral mutations aimed at slowing down encryption activity, which is achieved through \name's tiered architecture. The second is to provide a configurable balance between detection time and false positive rates, which is accomplished by requiring $K$ consecutive windows with detected malicious activity. Section~\ref{sec:det_time} elaborates on this trade-off. 

\subsection{Feature Analysis}
\label{sec:feature_analysis}

\name leverages seven distinctive file-level features that accurately describe the behavioral profile of files: read entropy, write entropy, write ratio, read ratio, number of processes, number of operations, and read/write ratio. We carefully selected these features following an analysis of a dataset containing both benign and ransomware processes to validate our initial intuition behind the choice. Our experimental evaluation demonstrates the significant challenge ransomware faces in modifying its behavior to successfully evade all of these features, given the contrastive nature between different features. 

\subsubsection{Read - Write Data Mismatch.}
The type of data read and written to a file is an important indicator of whether the file is seeing benign or malicious activity. 
For benign processes, the distribution of data read from a file and the distribution of data written to it are typically consistent. Conversely, ransomware tends to read non-encrypted data from files and write back encrypted data, exhibiting entirely distinct distributions. A sufficient approximation to capture this discrepancy can be obtained using entropy as a proxy for the distribution of data, as illustrated in Figures~\ref{fig:hist_rw_ent} and~\ref{fig:cdf_rw_ent}. These figures respectively plot the distribution and the CDF of the ratio between read and write entropy of operations performed by benign and ransomware. As we can see, benign file activity results in a Gaussian-like distribution centered around $1.0$, indicating that the read and write distributions roughly match. Ransomware, on the other hand, display a distribution that concentrates in the interval $[0, 1]$, indicating higher entropy distribution for write operations compared to read operations. Similarly, the CDF highlights this same behavioral difference. While average operation entropy in isolation is not a robust metric~\cite{de_gaspari_reliable_2022}, we demonstrate that, by incorporating it in our contrastive design, it provides sufficient and reliable information to aid in classification. We further discuss this in Section~\ref{sec:resilienceAdaptive}.

\subsubsection{File Write Ratio.}
Typically, benign processes write (or overwrite) small chunks of files over time, rarely rewriting an entire file. From the standpoint of file behavioral profiles, this implies that within a short time window, the number of distinct bytes written by all processes acting on the file should be significantly smaller than the file size. Conversely, to increase the probability of receiving a ransom, ransomware needs to completely overwrite as many files as possible, as swiftly as possible. Consequently, in a time window where ransomware activity is present, most or all of the file content will be overwritten. This behavioral asymmetry also holds for multiprocess ransomware that distributes write operations among multiple processes~\cite{de_gaspari_evading_2022}. 
Although each individual ransomware process writes only a fraction of the overall file, from the standpoint of file-level behavioral profiles, the combined activity of all ransomware processes results in a high write ratio. 
The percentage of a file written in a time window is, therefore, a defining, robust feature that helps characterize operations performed by both traditional and evasive multiprocess ransomware.

\subsubsection{File Read Ratio.}
While it is uncommon for a benign process to entirely overwrite a file, it frequently occurs that a benign process reads a file in its entirety. This behavior is prevalent for various user files, such as text documents, spreadsheets, and many others. Indeed, both ransomware and benign processes exhibit similar behaviors concerning the percentage of a file read, making it appear a weak feature for classification. However, its significance becomes evident when paired with the percentage of a file written. 
The primary techniques employed by evasive multiprocess ransomware to reduce feature expression are load splitting and function splitting: an individual ransomware process is divided into multiple processes, and both the number of operations and their type (i.e., read, write), are distributed among the processes. Splitting functionality among processes implies that, in certain cases, ransomware processes perform solely read operations on files, while sharing the read data with multiple writer processes through some stealth communication channel~\cite{de_gaspari_evading_2022,zhou2023limits}. The writer processes then encrypt the data and overwrite the file at some later time. From the perspective of the file-based behavioral profile, this implies that there may be instances in which write operations of encrypted data occur within a time window, with no corresponding read operation.

Figures~\ref{fig:hist_rw_ratio} and~\ref{fig:cdf_rw_ratio} highlight the fundamental difference between ransomware and benign processes behavior concerning file read/write ratio mismatch. We observe a substantial distinction between ransomware activity and benign process behavior both in the read/write ratio of files, thereby supporting our hypothesis that this feature offers valuable information for distinguishing between benign and ransomware activity in the system.

\subsubsection{Number of Processes Reading or Writing the File.}
Generally, in a time window with no ransomware activity, user files are accessed by a limited number of processes, often just one. This behavior also applies to certain traditional ransomware, where a single process reads and writes a file entirely. However, some traditional ransomware and all evasive multiprocess ransomware rely on the coordinated action of multiple processes targeting the same file, either to accelerate encryption or to evade detection. To imitate the behavior of benign processes, evasive multiprocess ransomware impose strict limits on the number of operations that each individual process is allowed to perform, as well as on the size of such operations~\cite{de_gaspari_evading_2022,zhou2023limits}. This is achieved by spawning multiple distinct processes to read and write the file in rapid succession, thereby fully encrypting user files while satisfying these constraints. Consequently, the number of processes performing read or write operations on a file within a time window serves as a crucial feature for detecting both traditional and multiprocess ransomware. 

Figures~\ref{fig:bar_proc_num} and~\ref{fig:cdf_proc_num} illustrate the distribution and CDF of the number of processes acting on a file in a 1-second time window. We observe the inherent difference between ransomware and benign process behavior. The plots indicate that, in nearly all cases, only a single benign process accesses a file within a time window. While this behavior is also observed for some types of traditional ransomware, a clear distinction emerges for the remainder and for multiprocess ransomware.

\subsubsection{Number of File Operations.}
The number of read and write operations that a file undergoes within a given time window offers essential context to weigh the significance of the other features. Benign processes commonly execute occasional file operations of relatively small size over their lifetime, leading to file-based behavioral profiles characterized by a low number of operations within each time window. This is apparent from Figures~\ref{fig:hist_op_num} and~\ref{fig:cdf_op_num}, which depict the distribution and CDF of the number of operations received by a file over 1-second intervals. Certain traditional ransomware favor using a limited number of large read and write operations during the encryption process, likely aiming to maximize I/O performance. Conversely, the majority of traditional ransomware utilize numerous small operations on the file, presumably to minimize the likelihood of detection. Similarly, in an effort to closely imitate the behavior of benign processes, evasive multiprocess ransomware imposes restrictions on the number of operations performed by each individual ransomware process and on the size of each operation~\cite{de_gaspari_evading_2022,zhou2023limits}. The restriction imposed on individual operation size forces evasive multiprocess ransomware to use a greater overall number of read/write operations, distributing them across the various ransomware processes. In terms of file-based behavioral profiles, this behavior leads to an elevated number of file operations within a time window characterized by malicious activity, in contrast to one with only benign activity. These observations are validated by Figures~\ref{fig:hist_op_num} and~\ref{fig:cdf_op_num}, which highlight the propensity of ransomware to execute a substantially higher number of operations on a file within a given time window.
Lastly, the number of operations on the file complements the number of processes feature, introducing another valuable dimension for the classifier to consider.

\subsection{Adaptive Attack Resilience}\label{sec:resilienceAdaptive}
When machine learning approaches are deployed in adversarial contexts, it is important to account for an \emph{adaptive} adversary in the threat model, who alters their attack to evade the specific detector encountered. This section elaborates on the various types of adaptive attacks that may be crafted against \name, and outlines why our detector is resilient against them.

\subsubsection{Evading the Read - Write Data Mismatch.}
\name uses average operation entropy to identify mismatches between the distribution of data read from and written to a file. However, adversaries can easily manipulate the average entropy of operations, as demonstrated in literature~\cite{naked_sun_acns}. Considering that the average entropy is computed across all operations within a time window, an adaptive ransomware attack simply requires injecting dummy low-entropy operations among the high-entropy encryption operations to evade this feature. These low-entropy operations can be executed by a single process for traditional ransomware or distributed across multiple processes for evasive multiprocess ransomware. 

As we demonstrate in Section~\ref{sec:adaptive_eval}, \name is inherently resilient to this family of adaptive attacks. Indeed, arbitrary manipulation of the average entropy requires executing additional operations on the file within the same time window. While these extra operations achieve the intended goal of reducing average entropy, they also inadvertently increase the number of operations feature due to \name's contrastive design, ultimately leading to detection.

\subsubsection{Evading File Read and Write Ratios Features.} 
The file read and file write ratio features are directly correlated to the encryption speed of ransomware. The only method adaptive ransomware can adopt to alter these features is slowing down its encryption activity, leading to only a small portion of the file being read/written within a time window. However, this poses a significant challenge for ransomware, as its primary objective is to encrypt as many files as swiftly as possible before detection. If the ransomware were to slow down the encryption process excessively, it would render itself ineffective~\cite{kirda_unveil,kirda_redemption}. 

We demonstrate in Section~\ref{sec:adaptive_eval} that, due to its tiered architecture, \name effectively detects this family of adaptive ransomware, even when it employs substantial slowdowns of up to $80\%$.

\subsubsection{Evading Number of Processes Reading or Writing the File Feature.}
The number of processes feature is primarily addressed at detecting evasive multiprocess ransomware, which uses multiple independent processes to encrypt files. To evade this feature, adaptive ransomware would be required to reduce the number of processes operating on a file within a time window, thereby compromising its ability to distribute tasks across processes and imitate benign behavior. In turn, this leads the file-based profile of evasive multiprocess ransomware to resemble that of traditional ransomware. As shown in the experimental evaluation, \name excels at detecting this type of ransomware.

\subsubsection{Evading Number of Operations Feature.}
The number of operations feature primarily facilitates the detection of ransomware that attempts to minimize the size of its read/write operations. Both traditional and evasive multiprocess ransomware commonly adopt this strategy to make it more challenging for traditional detectors to identify their activity. Decreasing the number of operations within a time window requires a proportional increase in the size of individual reads/writes, thereby simplifying the detection task for traditional detectors such as antivirus.

Nonetheless, we demonstrate in Section~\ref{sec:adaptive_eval} that \name is robust against this family of adaptive ransomware.

\section{Experimental Setup}\label{sec:expsetup}
\subsection{Dataset}
\label{sec:dataset}
\subsubsection{Low-Level File Operations Data.}\label{sec:irplogs_dataset}
We utilize low-level file operation data (IRP logs) from three distinct types of ransomware: traditional, evasive multiprocess, and adaptive. Traditional ransomware includes operation data from 383 samples gathered in previous works~\cite{continella_shieldfs:_2016,mehnaz_rwguard} --- called Traditional Classic --- along with data from 43 additional samples we collected from recent families (2023-2025) --- called Traditional Current. Details on the collection process are provided in Appendix~\ref{sec:ransomware_collection}. For evasive multiprocess ransomware, we incorporate operation data from~\cite{naked_sun_acns}, covering all proposed evasive configurations. Adaptive ransomware operation data was generated by modifying the behavior of the evasive multiprocess ransomware to manipulate the features used by \name for detection, as detailed in Sections~\ref{sec:adaptive_ransomware_generation}. Finally, for benign processes, we use operation data from~\cite{continella_shieldfs:_2016}, which includes extensive user activity recorded on 11 different machines across a diverse set of applications, including office suites, development tools such as Visual Studio, and archiving programs like WinRar. In total, the dataset comprises 476 ransomware samples and over 2000 benign applications. Full details are provided in Appendix~\ref{sec:dataset_description}.

\subsubsection{Train and Test Datasets}\label{sec:features_dataset}\name uses an ensemble multi-tier architecture, where each tier's classifier monitors file activity occurring within a specific time window. Our implementation uses the following time windows for each tier: $[0.25, 0.5, 1, 2, 4]$ seconds.
From the low-level file operations data, we compute \name features for each file and time window, deriving four separate datasets: (1) benign, (2) traditional ransomware, (3) evasive multiprocess ransomware, and (4) adaptive ransomware. We split the first three datasets in train and test based on a $70\%-30\%$ split and train \name on the three combined training datasets. The adaptive ransomware dataset is only used as a test set to assess the resilience of \name to adaptive attacks and not for training, with the exception of two R/W variants that are used to train \name on the relation between entropy and number of files operations, as detailed in Section~\ref{sec:adaptive_ransomware_generation}. Unless explicitly stated otherwise, all reported results are based on this train-test split.

\subsection{Adaptive Ransomware}\label{sec:adaptive_ransomware_generation}
We devise three new families of \textit{adaptive ransomware} specifically engineered to evade \name: (1) adaptive R/W entropy, (2) adaptive operation number, and (3) adaptive R/W ratio. These adaptive ransomware are enhanced versions of the mimicry evasive multiprocess ransomware~\cite{de_gaspari_evading_2022}, refined to adversarially manipulate one or more of \name's detection features. (1) The adaptive R/W entropy ransomware adjusts the read and write entropy features by injecting artificial, low-entropy read and write operations, thereby modulating the average entropy of operations within a given time window. We generate six variants of this adaptive family, with entropy reductions ranging from $2$x up to $8$x. We train \name on the first two variants to help the classifiers learn the relationship between entropy and the number of operations. The remaining four variants, which remain unseen during training, are used to evaluate \name's ability to generalize and detect previously unseen evasive strategies. (2) The adaptive operation number ransomware alters the number of operations executed within a time window by increasing the size of each individual operation. We developed nine variants of this family, reaching up to $90\%$ reduction in the number of operations. All variants are excluded from training. (3) The adaptive R/W ratio ransomware modifies the file read and write ratios within a specific window by slowing down the ransomware encryption activity. We generate six variants for this family, with up to $80\%$ slowdown in the encryption speed. All variants are unseen by the classifier during training. 

\section{Evaluation}\label{sec:evaluation}

This section evaluates the performance of \name. We show that \name consistently detects different types of ransomware with low false positive rates, i.e., without impacting benign activity in the system. Moreover, we demonstrate that the fast detection time of our approach minimizes file loss rate, preventing widespread damage to the system. We evaluate \name on both \textit{unseen} ransomware and \textit{unseen adaptive} ransomware, which is crafted specifically to evade our detector. In both cases, we prove the generalization ability of \name and the robustness of our design. Lastly, we analyze the explainability of our classifier and provide insights on how \name distinguishes between benign and ransomware classes.

\subsection{Metrics}
\label{sec:setup_metrics}
As \name is a \textit{file-based} detector, utilizing performance metrics based on the ratio of detected ransomware processes would be misleading. In each time window, \name constructs a behavioral profile for every opened user file in the system, and classification is conducted on a per-file basis within the window. Consequently, when evaluating \name's performance, our focus lies on the ratio of detected files exhibiting malicious activity, rather than the ratio of detected ransomware instances. Evaluating \name based on ransomware detection rate would trivialize the detection task: as long as malicious ransomware activity on even a single file is detected, the detection performance would register as $100\%$. This metric not only lacks significance but also misrepresents the effectiveness of the defense. For instance, a ransomware that successfully encrypts most of the file system but is only detected during the encryption of the last few files would still be counted as a successful detection. 

To better represent the effectiveness of our approach, we assess \name detection rate based on the number of files targeted by the ransomware that are successfully identified. We define True Positives (TP), True Negatives (TN), False Positives (FP), and False Negatives (FN) as follows.
\begin{itemize}
    \item TP: number of individual files exhibiting ransomware activity classified as malicious by \name.
    \item TN: number of individual files without ransomware activity classified as benign by \name.
    \item FP: number of individual files without ransomware activity classified as malicious by \name.
    \item FN: number of individual files exhibiting ransomware activity classified as benign by \name.
\end{itemize}

Given the above definitions, we use the following standard performance metrics in our evaluation: True Positive Rate (TPR), True Negative Rate (TNR), accuracy, precision, recall, and F1 Score (F1). 

When discussing our results, we denote the overall performance of the ensemble of \name models as \textit{stacked}, while the performance of each individual \name classifier is referred to by the duration of its corresponding window (e.g., 0.25sec classifier is the classifier using 0.25sec of activity on a file to make a prediction).

\begin{figure}
    \centering
    \includegraphics[width=0.80\columnwidth]{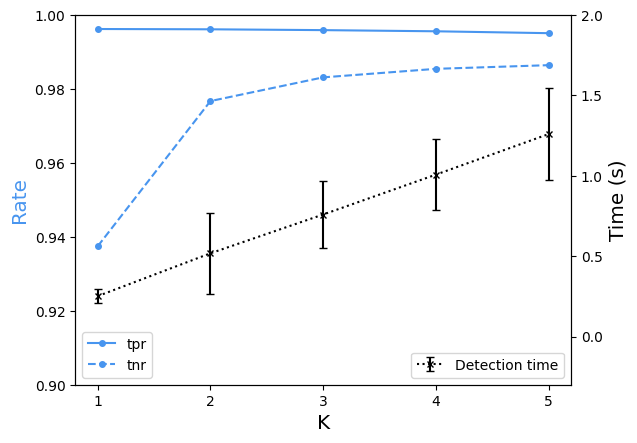}
    \caption{True positive rate, true negative rate, and average detection time with standard deviation for varying K.}
    \label{fig:detection_time_and_detection_rates}
\end{figure}

\subsection{Detection Time}
\label{sec:det_time}

\begin{figure}
    \centering
    \includegraphics[width=0.75\columnwidth]{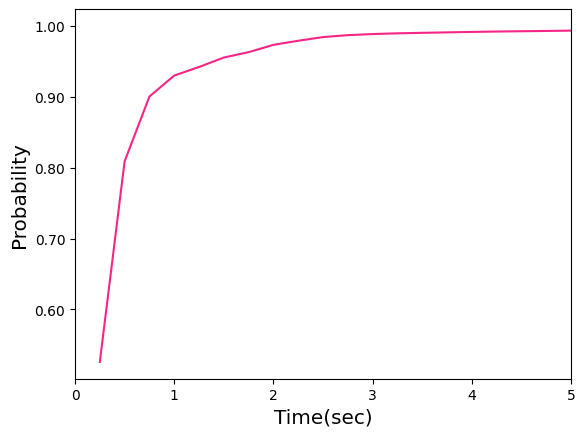}
    \caption{Cumulative Distribution Function of total ransomware activity duration on each file.}
    \label{fig:cdf_rans_dur}
\end{figure}

In Figure~\ref{fig:detection_time_and_detection_rates}, we report TPR, TNR, and the time to detect the ransomware activity in the system by \name for varying numbers of $K$. As mentioned in Section~\ref{sec:ourapproach}, $K$ indicates the number of consecutive windows where malicious activity is detected needed for \name to classify the operations as malicious. Different values of $K$ provide a trade-off to balance between detection time and false positive rates. Figure~\ref{fig:detection_time_and_detection_rates} shows that increasing the value of $K$ leads to a corresponding increase in $TNR$, indicating better end-user experience due to the decreased occurrence of benign processes flagged as ransomware. 
This is expected, as the likelihood of $K$ consecutive misclassifications is inversely proportional to the value of $K$. Conversely, the average detection time is directly proportional to $K$. Figure~\ref{fig:detection_time_and_detection_rates} shows the proportionality relation is linear, with the average detection time being almost exactly equal to the minimum window size times $K$. This is due to the high TPR achieved by the 0.25sec window classifier, as we discuss later in Section~\ref{sec:detection_eval}. Finally, the overall TPR is nearly unaffected by $K$, decreasing by ${\sim}0.4$ percentage points between $K=1$ and $K=5$. The motivation behind this behavior is twofold. First, the TPR of \name is consistently high across all windows with malicious activity, meaning that typically every window with ransomware activity is correctly flagged by \name. Therefore, given $K$ windows with malicious activity, the probability that at least one is flagged as benign is low. Second, as illustrated in Figure~\ref{fig:cdf_rans_dur}, the total duration of ransomware activity on a single file lasts less than 1 second for ${\sim}95\%$ of all files. This behavior effectively puts an upper bound on the number of maximum windows $k$ used for detection. When there are $n<K$ consecutive windows of activity on a given file, \name will attempt classification using the $n$ available windows, regardless of $K$.
Therefore, increasing $K$ beyond a certain value impacts the classification only on a very small percentage of files.

Based on this empirical evaluation, we found that $K=2$ provides the best trade-off between detection time, TNR, and TPR. For $K=2$, \name detects over 99\% of the ransomware activity within $0.52(\pm0.25)sec$ with low FPR, reducing the probability of significant data loss. Furthermore, as we discuss in Section~\ref{sec:discussion}, the fast detection rate of \name enables the adoption of near-zero overhead data recovery approaches, effectively nullifying data loss. Henceforth, unless explicitly stated otherwise, all subsequent experiments are presented for $K=2$.

\subsection{Ransomware Detection Performance}\label{sec:detection_eval}

\begin{table*}[t]
\small
\caption{Performance analysis of \name against traditional and multiprocess ransomware attacks.}
\centering
\begin{tabular}{c | cccccccc}
\toprule
 & Benign & Trad. Classic & Trad. Current & Multiproc. & \multicolumn{4}{c}{Overall}\\
\toprule
{Window} & $TNR$ & $TPR_{T}$ & $TPR_{C}$ & $TPR_{M}$ & Acc. & Prec. & Recall & F1 \\ 
\cmidrule(l){1-9}
0.25sec & 98.34 & 97.99 & 99.65 & 99.93 & 98.76 & 99.17 & 98.97 & 99.07 \\ 
0.5sec  & 98.38 & 98.50 & 99.66 & 99.95 & 98.95 & 99.19 & 99.24 & 99.21 \\ 
1sec    & 98.33 & 98.57 & 99.66 & 99.96 & 98.96 & 99.17 & 99.27 & 99.22 \\ 
2sec    & 98.46 & 98.97 & 99.66 & 99.95 & 99.13 & 99.23 & 99.46 & 99.35 \\ 
4sec    & 98.91 & 99.09 & 99.67 & 99.95 & 99.31 & 99.45 & 99.52 & 99.49 \\
\midrule
stacked & 97.25 & 99.23 & 99.70 & 99.98 & 98.82 & 98.63 & 99.61 & 99.12 \\ 
\end{tabular}
\label{tab:performance_analysis}
\end{table*}

\subsubsection{Traditional Ransomware}\label{sec:traditional_eval}
Traditional ransomware comprises ransomware families that are used in real-world attacks and can be found in the wild. These families of ransomware typically exhibit low degrees of parallelism and spawn few processes to encrypt the target system and perform other malware-related activities~\cite{naked_sun_acns}. Traditional ransomware sometimes employ multiple processes to speed up the encryption of the target system, but do not perform complex balancing of operations between the various processes to avoid detection~\cite{de_gaspari_evading_2022}.

Table~\ref{tab:performance_analysis} reports the performance of \name on the traditional ransomware test set, which includes Traditional Classic ($TPR_T$) and Current ($TPR_C$). The true positive rates value for all the considered time windows is $\geq98\%$, reaching over $99\%$ on the $4sec$ window. This means that all individual \name classifiers can accurately pinpoint traditional ransomware activity with less than 2\% of false negatives. Similarly, the TNR ranges between $98\%$ and $99\%$ across all windows, highlighting the ability of \name to effectively detect ransomware without compromising user experience. Finally, the overall stacked performance of \name, which aggregates the predictions of each individual classifier to take a decision, reaches over $99\%$ TPR. The improved TPR of stacked \name compared to the $4sec$ window classifier derives from the independent nature of the errors made by each individual classifier, which results in higher overall detection rate. The performance analysis in Table~\ref{tab:performance_analysis} emphasizes the ability of \name to generalize well across ransomware families, including very recent (2022-25) samples. This characteristic of \name is fundamental in ensuring rapid detection of ransomware activity, discussed in Section~\ref{sec:det_time}, which ensures low data loss rate and the compatibility of \name with near-zero overhead data protection techniques.

\subsubsection{Evasive Multiprocess Ransomware}\label{sec:evasive_eval}
Evasive multiprocess ransomware families are characterized by their extensive use of parallelism, relying on multiple processes to encrypt the target system. These processes collaborate in a stealthy manner in order to perform complex balancing of the ransomware operations between multiple processes to avoid detection~\cite{de_gaspari_evading_2022,zhou2023limits}. Some of these families leverage this covert inter-process collaboration to closely mimic benign process behavior~\cite{naked_sun_acns}, entirely evading detection by traditional process-based behavioral detectors~\cite{continella_shieldfs:_2016,mehnaz_rwguard,kirda_redemption,kirda_unveil}. 

Table~\ref{tab:performance_analysis} reports the performance of \name on the evasive multiprocess ransomware test set.  
The true positive rate ($TPR_M$) is largely above $99\%$ for all of the time windows, which confirms our hypothesis that \name's file-based behavioral profiling is resilient to this type of evasive multiprocess ransomware, contrary to prior works.
We also note that there is essentially no performance variation between the different \name classifiers. While initially perplexing, this phenomenon can be explained upon closer analysis of the behavior of the multiprocess ransomware. By distributing the workload across multiple processes, the activity of the ransomware on each individual file typically lasts less than $0.5sec$ --- the minimum time required for detection by \name with $K=2$. Consequently, all \name classifiers see the full ransomware activity on the file, which explains the negligible performance variations between different windows. Furthermore, as elaborated later in Section~\ref{sec:shap_evaluation}, the variability in behavior among the various processes of multiprocess ransomware is minimal compared to traditional ransomware, which aids in clarifying the near-perfect detection performance.

\subsection{Adaptive Detection Performance}\label{sec:adaptive_eval}

\begin{figure*}[t]
    \centering
    \begin{subfigure}{.32\textwidth}
        \includegraphics[width=\columnwidth]{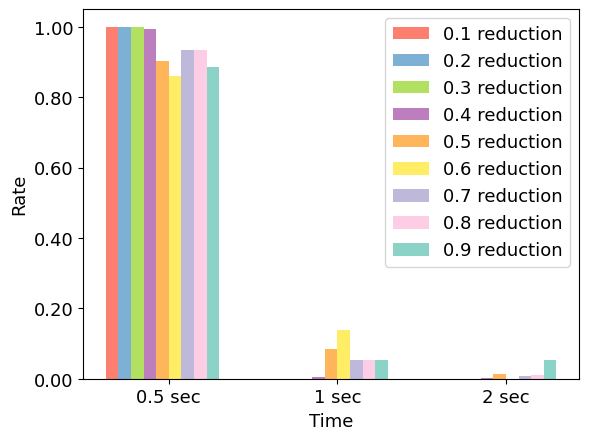}
        \caption{Detection rate of adaptive ransomware relying on reducing the number of operations for different reduction factors, K=2.}
        \label{fig:detection_rate_ad_opnum}
    \end{subfigure}  
    \hfill
    \begin{subfigure}{.32\textwidth}
        \centering
        \includegraphics[width=\columnwidth]{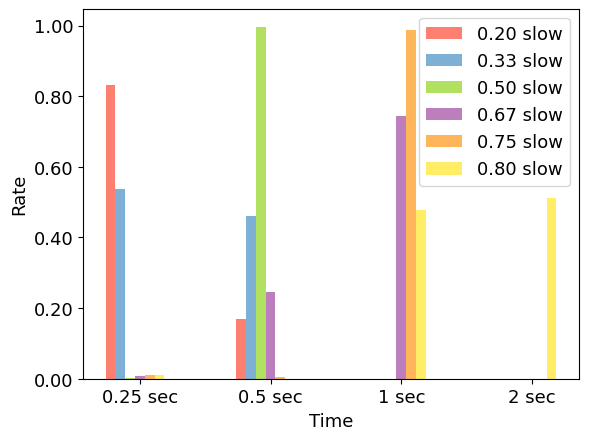}
        \caption{Detection rate of adaptive ransomware relying on reducing the read/write frequency for different slowdown factors, K=1.}
        \label{fig:detection_rate_ad_rw_k1}
    \end{subfigure}
    \hfill
	\begin{subfigure}{.32\textwidth}
        \centering
        \includegraphics[width=\columnwidth]{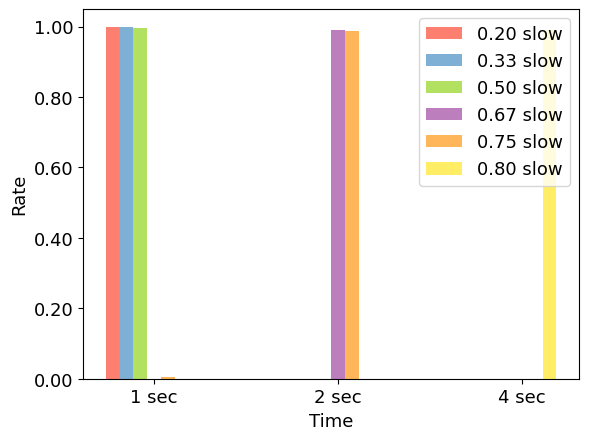}
        \caption{Detection rate of adaptive ransomware relying on reducing the read/write frequency for different slowdown factors, K=2.}
        \label{fig:detection_rate_ad_rw_k2}
    \end{subfigure}
    \caption{Adaptive ransomware detection rate, for different adaptive approaches and different configurations.} 
    \label{fig:detection_plots_adaptive}
\end{figure*}

\begin{table}[t]
\caption{Average stacked detection rate of \name for different adaptive ransomware families.}
\centering
\begin{tabular}{c | c | c | c}
\toprule
Adaptive Type & $TPR$ & \makecell[c]{Mean Det. Time\\(sec)} & \makecell[c]{Det. Time\\Std. Dev.} \\ 
\midrule
R/W Entropy &	99.93   &   0.50	&    0.02 \\
R/W Ratio   &   99.94   &   1.82	&    1.07 \\
Op. Num.    &	98.16   &   0.56	&    0.45 \\
\end{tabular}
\label{tab:ad_performance}
\end{table}

We assess the effectiveness of \name's contrastive design by evaluating its performance against three adaptive ransomware families specifically engineered to evade detection (see Section~\ref{sec:adaptive_ransomware_generation}).
Table~\ref{tab:ad_performance} illustrates the stacked detection rate for these adaptive families. Notably, the TPR remains largely unaffected by the behavioral modifications of the different adaptive families. Specifically, adaptive ransomware targeting the read-write entropy features are promptly detected by \name within $0.5sec$, which is the minimum detection time when $K=2$. Similarly, adaptive ransomware altering read- and write-ratio behaviors are also consistently detected. However, the mean detection time for this adaptive family is significantly larger at $1.82\pm1.07$sec. These findings are expected, as the reduction in the R/W ratio feature is achieved by aggressively slowing down encryption activity (up to $80\%$ slowdown, as outlined in Section~\ref{sec:adaptive_ransomware_generation}). Nevertheless, detection performance remains excellent, as the classifiers leveraging long-term windows effectively identify 
the slow encryption activity. The sole adaptive family achieving limited success in evading \name are the adaptive operation number ransomware. These ransomware reduce the number of read and write operations by increasing the size of each individual operation. However, as highlighted in Table~\ref{tab:ad_performance}, \name continues to deliver excellent detection performance against this adaptive family. Furthermore, since this evasion strategy entails reducing the number of operations while increasing their size, attempting to evade \name using this method renders the attack easily detectable through process-based detection~\cite{de_gaspari_evading_2022}. Finally, we also evaluated \name against compatible combinations of evasive strategies (r/w entropy + r/w ratio, r/w ratio + op. num.), obtaining similar detection results.

In Figure~\ref{fig:detection_plots_adaptive}, we further study the detection rate distribution across \name's classifiers for the different adaptive families. We exclude from this analysis r/w entropy adaptive family, as it is always detected by the short-term $0.25sec$ classifier. Figure~\ref{fig:detection_rate_ad_opnum} plots the detection rate against adaptive ransomware that manipulates the number of operations feature. From the figure, we notice that the vast majority of variants ($>90\%$) are detected within $0.5sec$ by the short-term $0.25sec$ window when $K=2$. These results are expected, as this adaptive family does not attempt to slow down encryption activity, and therefore we expect a detection time in line with the overall mean detection time. Figures~\ref{fig:detection_rate_ad_rw_k1} and~\ref{fig:detection_rate_ad_rw_k2} depict the detection rate distribution for the adaptive R/W ratio family of ransomware for $K=1$ and $K=2$ respectively. 
The detection distribution among the \name classifiers skews towards longer-term classifiers as the slowdown ratio increases, as expected. Furthermore, interesting insights can be derived by comparing the detection distribution for $K=1$ and $K=2$. Initially, one might expect the detection distribution for $K=2$ to closely resemble that of $K=1$, albeit with double the detection time. However, we observe that this expectation is incorrect. In reality, the minimum detection time for adaptive R/W ransomware increases fourfold between the two settings, while the maximum detection time increases twofold, as initially anticipated. This behavior is explained by the weaker detection performance of \name's short-term classifiers on individual windows of ransomware activity for this adaptive family. While the short-term term classifiers effectively detect at least one window of malicious activity (sufficient for detection when $K=1$), they fail to detect two consecutive windows when the ransomware activity slows down. This ultimately leads to a bias in the detection distribution towards long-term classifiers, and, therefore, longer detection times.

\subsection{Feature contribution to decision-making}\label{sec:shap_evaluation}
This section assesses the explainability of \name and the contributions of each of feature to the final classification outcome. We leverage the SHapley Additive exPlanations (SHAP)~\cite{shappaper} technique, a widely adopted method to interpret model predictions. 
Features with positive SHAP values influence the prediction towards the positive class (and vice-versa), while the magnitude of each individual feature indicates the strength of its contribution to the final prediction. As a reminder, \name is a \textit{file-based} behavioral classifier. Therefore, it is understood that the ensuing discussion on feature importance and behavioral profiles is conducted from the perspective of activity on individual files.

\begin{figure*}[t]
    \centering        
	\begin{subfigure}{.32\textwidth}
            \centering
            \includegraphics[width=\columnwidth]{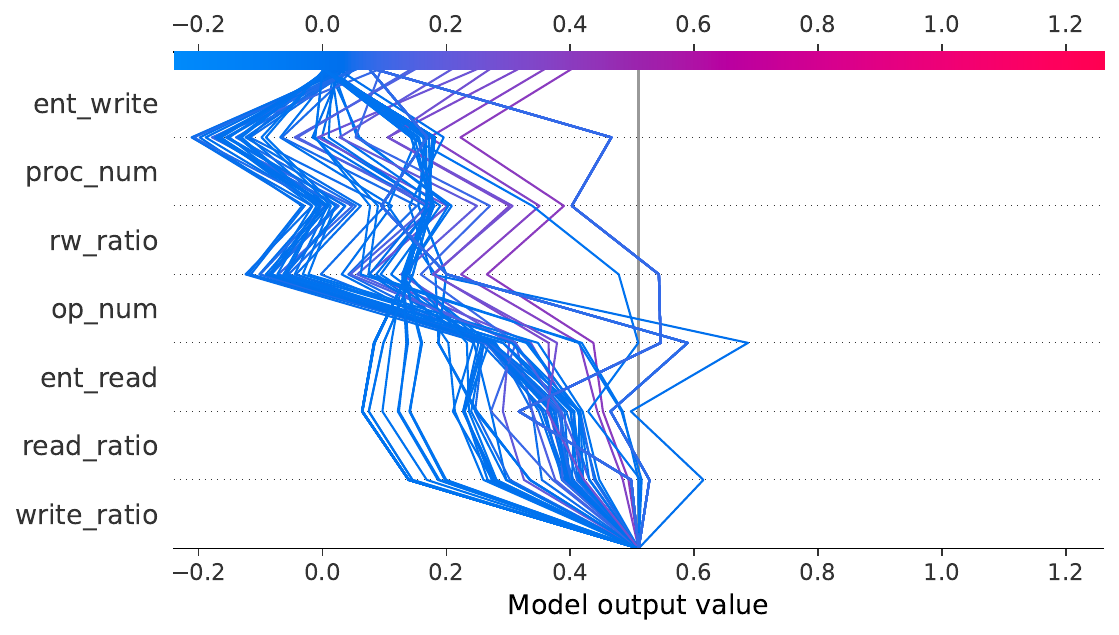}
             \caption{SHAP values on 1 second window, benign process features.}
             \label{fig:shap_1_benign}
        \end{subfigure}
        \hfill
	\begin{subfigure}{.32\textwidth}
            \centering
             \includegraphics[width=\columnwidth]{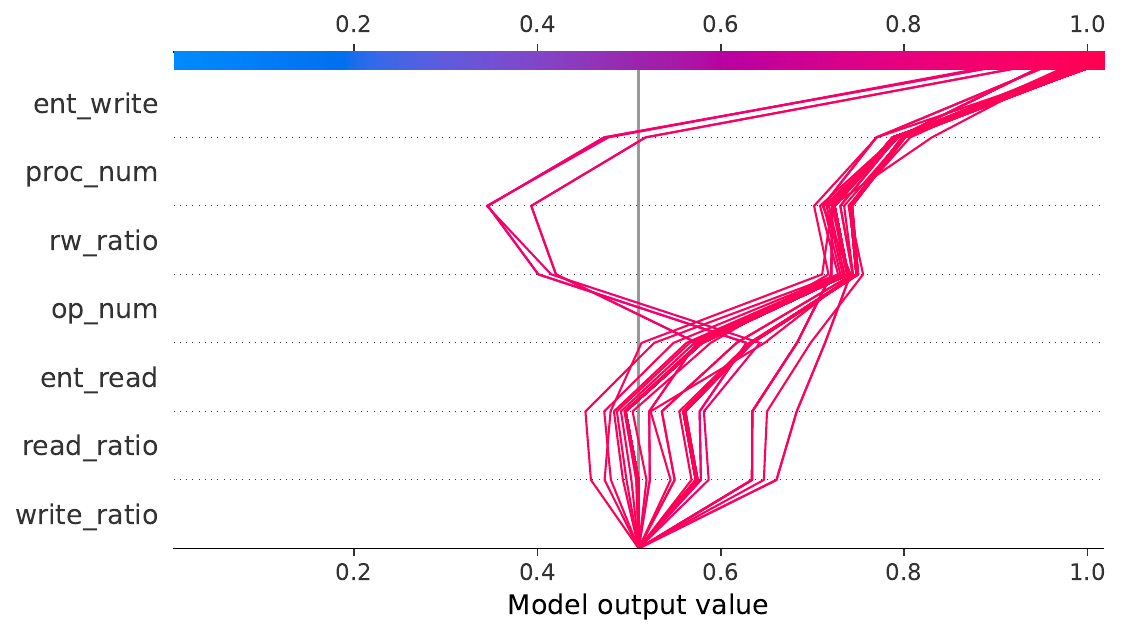}
             \caption{SHAP values on 1 second window, traditional ransomware.}
             \label{fig:shap_1_trad}
        \end{subfigure}
        \hfill
	\begin{subfigure}{.32\textwidth}
            \centering
            \includegraphics[width=\columnwidth]{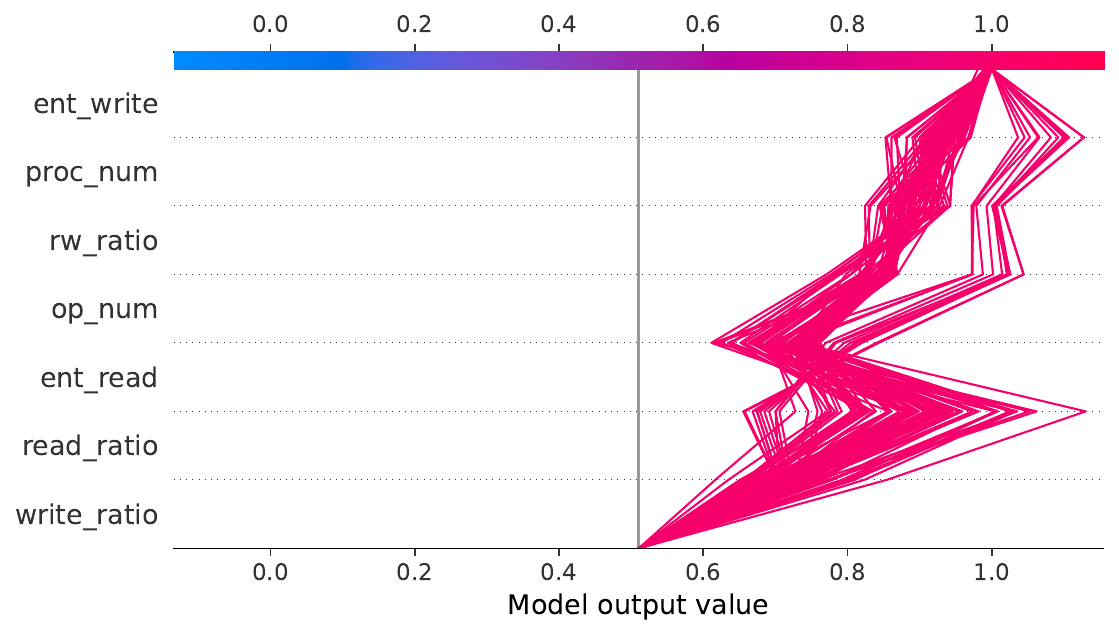}
             \caption{SHAP values on 1 second window, multiprocess ransomware.}
             \label{fig:shap_1_cerb}
        \end{subfigure}
	\begin{subfigure}{.32\textwidth}
            \centering
             \includegraphics[width=\columnwidth]{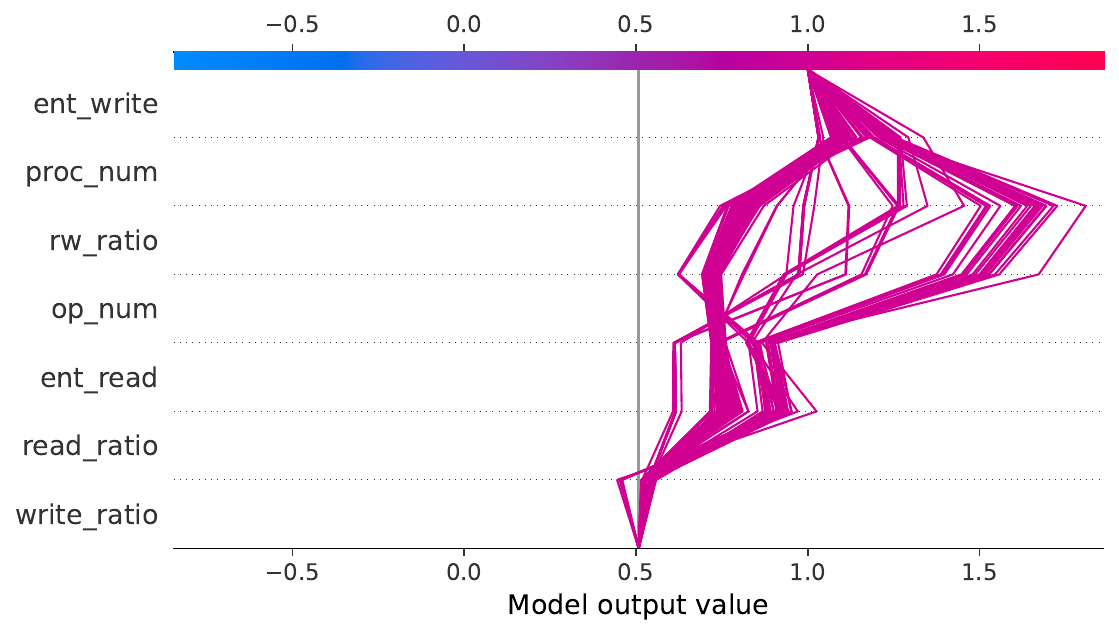}
             \caption{SHAP values on 1 second window, adaptive entropy ransomware.}
             \label{fig:shap_1_adaptive_ent}
        \end{subfigure}
        \hfill  
	\begin{subfigure}{.32\textwidth}
            \centering
            \includegraphics[width=\columnwidth]{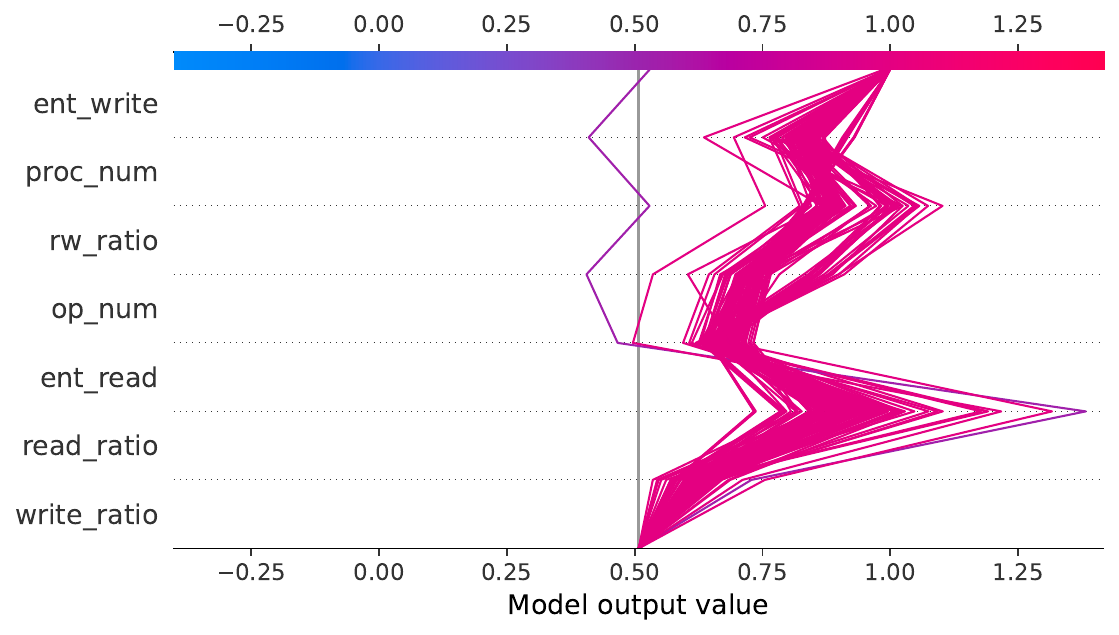}
             \caption{SHAP values on 1 second window, R/W ratio adaptive ransomware.}
             \label{fig:shap_1_ad_rw}
        \end{subfigure}
        \hfill
	\begin{subfigure}{.32\textwidth}
            \centering
            \includegraphics[width=\columnwidth]{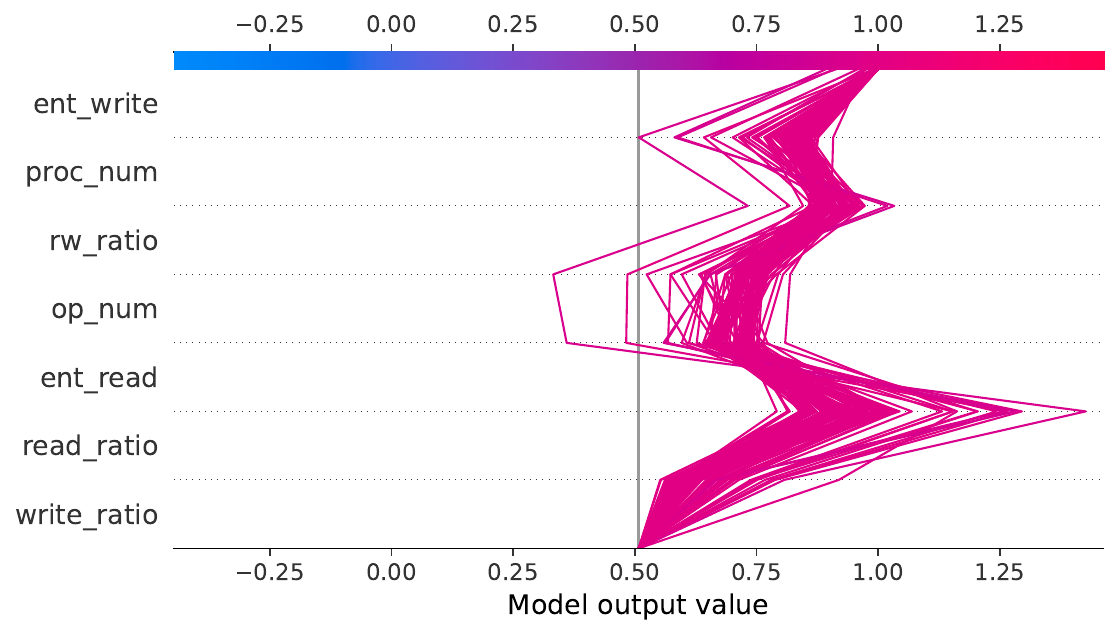}
             \caption{SHAP values on 1 second window, Operation Number adaptive ransomware.}
             \label{fig:shap_1_adaptive_opnum_5th}
        \end{subfigure}  
    \caption{SHAP-based analysis of the \name features on the detection of benign and different ransomware types considered (i.e., traditional, multiprocess and adaptive).} 
    \label{fig:feature_importance_plots}
\end{figure*}

Figure~\ref{fig:feature_importance_plots} presents the SHAP plots for the $1sec$ window \name classifier, highlighting the importance of each selected feature to the final classification and how feature importance changes for different types of ransomware. Each plot presents the SHAP values of \name's features for 100 randomly sampled data points. 
Figures~\ref{fig:shap_1_benign},~\ref{fig:shap_1_trad}, and~\ref{fig:shap_1_cerb} display the individual contributions of the \name features to the prediction of benign, traditional ransomware and multiprocess ransomware classes for the one-second window model. An immediate observation is the significance of the write ratio feature as a distinguisher between benign and ransomware. This finding is expected, as benign applications typically perform partial writes to files, while ransomware aims to fully rewrite user files as part of its malicious activity. A similar pattern emerges with the read ratio feature, where benign applications often read only portions of files, while ransomware tend to read files fully to encrypt them. Figure~\ref{fig:shap_1_trad}, however, reveals exceptions to this trend, indicating that some traditional ransomware variants also exhibit partial file read and write behaviors. We hypothesize that these ransomware variants prioritize speed over completeness of encryption, sacrificing full file encryption to achieve higher overall encryption speed. Other important features are operation number and process number, with ransomware typically exhibiting a notably higher number of operations and processes acting on files during the encryption activity. This characteristic is especially prominent for multiprocess ransomware.
Finally, the read and write entropy features play a crucial role in detection, with the write entropy feature having a particularly strong influence on the model's classification for ransomware. Conversely, the impact of these features on benign activity is ambiguous. This outcome is expected, as certain benign applications display read/write entropy profiles akin to ransomware, as previously outlined in Section~\ref{sec:feature_analysis}.

Figures~\ref{fig:shap_1_ad_rw} and~\ref{fig:shap_1_adaptive_opnum_5th} illustrate the SHAP plots for adaptive R/W ratio and adaptive operation number, respectively. We observe that these plots exhibit minimal variation when compared to the non-adaptive multiprocess ransomware depicted in Figure~\ref{fig:shap_1_cerb}. These findings are expected, as the behavioral changes introduced by these adaptive ransomware families have limited impact on individual features of \name. 
For adaptive R/W ransomware, the behavioral changes typically affect only short-term classifiers, while long-term classifiers still capture the complete malicious activity. Adaptive operation number ransomware primarily alter the number of operations executed during encryption activities, which is evident in Figure~\ref{fig:shap_1_adaptive_opnum_5th}, where the contribution of the operation number feature on the final classification is negligible. However, the remaining features still lead to detection. Lastly, Figure~\ref{fig:shap_1_adaptive_ent} clearly illustrates the influence of contrastive features such as entropy and number of operations on classification outcome. We observe that while the entropy features exhibit a minimal to negative impact on ransomware classification, the contribution of the operation number features shows a considerable increase. This behavior is expected, as adaptive entropy ransomware relies on dummy, low entropy operations to decrease the average read/write entropy and evade detection.

\subsection{Overhead Analysis}
\label{sec:overhead}

\subsubsection{Disk Activity Monitor overhead}
\begin{figure*}[t]
    \centering
    \begin{subfigure}{.32\textwidth}
        \includegraphics[width=\columnwidth]{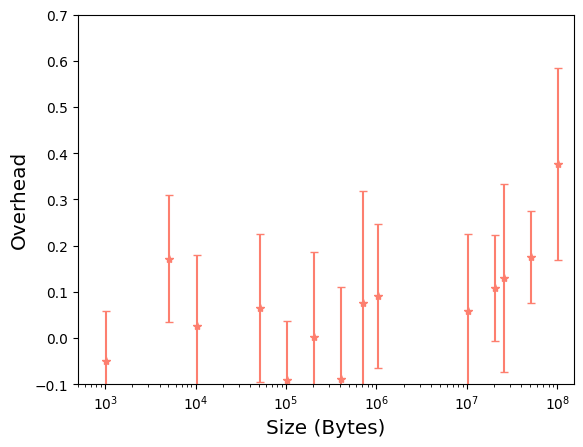}
        \caption{Overhead on file read operations under continuous I/O load.}
        \label{fig:read_overhead}
    \end{subfigure}  
    \hfill
    \begin{subfigure}{.32\textwidth}
        \centering
        \includegraphics[width=\columnwidth]{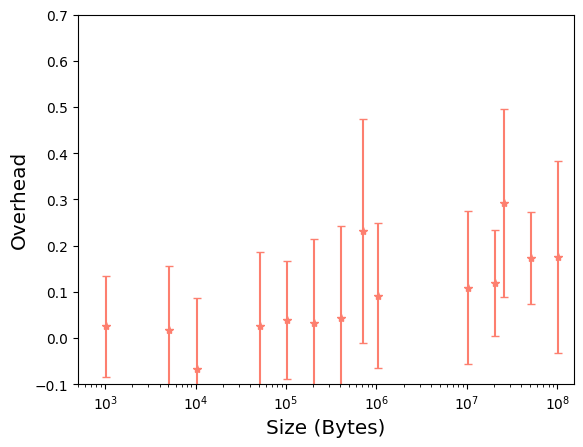}
        \caption{Overhead on file write operations under continuous I/O load.}
        \label{fig:write_overhead}
    \end{subfigure}
    \hfill
	\begin{subfigure}{.32\textwidth}
        \centering
        \includegraphics[width=\columnwidth]{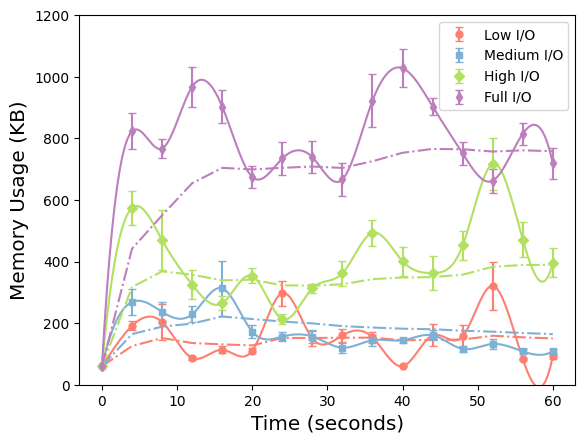}
        \caption{Memory overhead under different I/O conditions. Rolling averages as dashed lines.}
        \label{fig:memory_overhead}
    \end{subfigure}
    \caption{Overall system overhead introduced by \name. }
    \label{fig:all_overhead}
\end{figure*}

\name's Disk Activity Monitor module leverages an I/O Request Packets Logger driver (IRPLogger) to capture relevant file system activity~\cite{continella_shieldfs:_2016}. 
To evaluate the overhead introduced by the Disk Activity Monitor, we perform continuous read and write operations of different sizes ranging from 1KB to 100MB, simulating workloads on both small and large files, and measure the corresponding completion times.
Each operation was repeated $1,000$ times for reading and $1,000$ times for writing for each considered operation size, resulting in a total of $28,000$ file system operations. To avoid caching-related side effects, all operations were conducted on separate individual files. The tests were performed on a Windows 11 virtual machine equipped with an NVMe solid-state drive, 16 CPU cores, and 16GB of RAM. As shown in Figures~\ref{fig:read_overhead} and~\ref{fig:write_overhead}, \name's overhead during continuous I/O operations remains low, ranging from $0-20\%$ for all read and write operations, with an outlier reaching $\sim35\%$. On average, the overhead recorded across all operations was $6.76\%(\pm11.49)$. 

To further characterize the resource requirements of \name, we analyze its main memory and CPU usage under varying I/O loads. Figure~\ref{fig:memory_overhead} presents \name's main memory usage in KB as processes perform operations at random intervals, ranging from a few seconds for low I/O to continuous operations at full I/O. Memory usage exhibits a consistent pattern across all I/O loads, with rapid increases in the first few seconds of activity followed by fluctuations based on the number of requests and the size of the operations. The running average, represented by the dashed lines in the figure, stabilizes early in the execution and remains relatively constant throughout. This behavior aligns with \name's design, as it only stores file system operations within a maximum time window of 4 seconds, ensuring efficient memory usage and scalability even under high-intensity I/O workloads. Across our tests, we observe a negligible increase of $2.31(\pm1.09)$ percentage points in CPU usage when \name is active in the system.

Finally, we ran a small scale user study to assess whether there is any perceived usability degradation when using \name in day-to-day tasks. We asked 5 users to interact with two systems VMs: one running \name and one without. The users were allowed to freely use the system, but were asked to include browsing, document editing, software installation, and file transfers operations. After they completed their activity, we asked them to evaluate their experience on a scale from 1 to 5 with each system, where 1 corresponded to ``unusable system'' and 5 corresponded to ``perfectly usable''. Users reported an average of $4.8$ and $4.6$ for the system with and without \name, respectively. We attribute the perceived usability degradation on the system without \name to the use of virtualization.

\subsubsection{Sample prediction time}
\name's detectors leverage fully-connected DNN classifiers to identify ransomware activity. We evaluate the overhead of \name's Behavioral Detector module in predicting file activity by calling the \textit{predict} function on $1,000$ individual samples sequentially. The prediction function invokes the individual prediction functions of each tier's classifier, characterizing the overall time required to make a prediction across the full multi-tier architecture of \name.  Our tests indicate that \name's classification requires on average $0.0138(\pm0.0016)sec$ to predict a sample, introducing negligible impact on the system and \name's decision-making pipeline. Furthermore, while we tested sequential classification to evaluate the worst-case scenario, we stress that, in practice, classification operations can be parallelized in large batches thanks to the small size of \name's feature vectors (seven float32 features, or 224 bits each).

\section{Discussion and Limitations}\label{sec:discussion}
We identify three main limitations of \name: (1) potential data loss before detection, (2) coarse-grained detection granularity, and (3) potential susceptibility to advanced evasive ransomware.

\textbf{Data Loss.}
While \name demonstrates significant effectiveness in promptly identifying ransomware activity, there remains a potential risk of data loss before detection occurs. A detection time of $0.52(\pm0.25)$ seconds, as achieved by \name, translates to an average of 19 files being encrypted before detection by contemporary high-speed ransomware~\cite{splunk_rans_speed}. Advanced ransomware families such as LockBit and Babuk are capable of encrypting up to ${\sim}130$ files within the same time frame~\cite{splunk_rans_speed}, and future ransomware families will certainly keep improving.
While currently \name does not implement file recovery techniques, these can be implemented in the future to prevent any data loss. Given \name's rapid detection rate, transparent, near-zero overhead data recovery methods such as cache-based backups can be utilized~\cite{10017139}. 
Modern operating systems heavily rely on I/O buffering, which utilizes a page cache to temporarily store data in the main memory before transferring it to I/O devices. Portions of the page cache are periodically flushed at different intervals depending on the operating system: $1/8th$ of the cache every second for Windows~\cite{10.5555/2432156}, while Linux retains pages for up to 30 seconds by default. 
Since \name's detection rate is significantly faster than the cache flushing rate, it can be integrated with techniques that adjust cache flushing policies to prevent flagged malicious content from being written to disk. This approach would enable a near-zero overhead data recovery process~\cite{10017139}.

\textbf{Detection Granularity.}
\name detects ransomware activity on a per-file, per-window basis. This behavior inherently limits the detection granularity to the file level rather than the process level. When \name detects ransomware activity on a file in a time window, \textit{all} processes that interacted with the file during that window are classified as ransomware and suspended. This approach may include benign processes that wrote data to the file within the same window as the ransomware. We argue, however, that this is a minor limitation. First, operating systems prevent concurrent writes on the same file. Therefore, while the ransomware is encrypting a file, no benign write activity can happen. Second, since \name detection time is very low, even if benign activity happened in the same window as ransomware, on average, only $0.5sec$ of benign activity would be lost. While this might be acceptable for many types of usage, it might not be suitable in high-frequency I/O environments. A partial solution to this problem is to integrate process control and process analysis tools into \name. When ransomware activity is detected on a file, \name would freeze all processes interacting with it. Simultaneously, the user would receive a notification about the malicious activity and be prompted to confirm whether each identified process is legitimately accessing the file. Unverified processes would be terminated, and their file operations could be rolled back using the previously discussed cache-based recovery method. Such an approach would require the implementation of tools to help users identify processes and cannot be fully automated reliably. However, we argue that the benefits of promptly stopping ransomware activity outweigh the potential disadvantages.

\textbf{Advanced Evasive Ransomware.}
As ransomware evolves, future variants may adopt increasingly sophisticated evasion techniques. While \name is robust to several types of evasion attacks, a dedicated adversary may be able to still evade detection. For instance, ransomware could be designed to generate separate encryption processes for each file in the system, which would lead to high data loss. While creating so many new processes would be quickly detectable, sophisticated ransomware could employ a simplified version of this encryption method to maximize data loss. Advanced ransomware could coordinate its file activity with that of benign processes, reducing the effectiveness of some of \name's features such as read/write entropy, and potentially leading to evasion. Finally, given the reliance of \name on DNNs, future ransomware could utilize adversarial attacks leveraging carefully crafted noise to achieve evasion, such as adversarial examples~\cite{grosse_adversarial_2017}. While such attacks are complex and require advanced knowledge, they are not impossible and could lead to failed detection. Extending \name to consider further complex, coordinated adversarial attacks remains an open research direction.

\section{Conclusions}\label{sec:conclusions}
We presented \name, a novel approach to ransomware detection and mitigation. In contrast to existing detection methods, \name constructs behavioral profiles of files based on all the operations they receive within a specific time window. Thanks to its contrastive design approach, \name's detection remains robust against complex multiprocess ransomware and adaptive ransomware specifically engineered to evade detection.
Through a comprehensive evaluation, we demonstrated that \name is highly effective against all types of known ransomware attacks, unseen ransomware attacks, as well as evasive variants designed to bypass \name, underscoring the efficacy of our approach. Lastly, \name exhibits rapid ransomware activity detection, enabling the adoption of near-zero overhead data loss prevention techniques. We believe that these findings represent a significant advancement in the development of robust and resilient countermeasures against the escalating ransomware threat landscape.

\begin{acks}
This work was partially supported by SERICS (PE00000014) under MUR NRRP0 funded by the European Union and project AutoAD: Using Active Defense to Defeat Cyber Adversaries funded by Sapienza University of Rome 2022 (RG1221816C839BF9).
\end{acks}

\bibliographystyle{ACM-Reference-Format}
\bibliography{bibliography}

\appendix
\section{Additional Setup Details}
\subsection{Dataset Analysis}
Data presented in Figure~\ref{fig:feature_distribution_plots} is computed over benign, traditional ransomware, and evasive ransomware process operations performed over a one-second time window. To improve the presentation, we filtered the dataset by removing null read and write operations (which are also ignored by \name).

\subsection{Dataset description}
\label{sec:dataset_description}
The dataset used in our experiments comprises data from multiple sources~\cite{continella_shieldfs:_2016,mehnaz_rwguard,de_gaspari_evading_2022,zhou2023limits} and additional data gathered as part of this study. Table~\ref{tab:dataset_ransomware} lists the types of ransomware, ransomware families, and the number of samples for each family. The dataset includes five Traditional Classic ransomware families for a total of 383 samples, which are gathered from previous works~\cite{continella_shieldfs:_2016}. We additionally gathered 43 samples from five different ransomware families released between 2023 and 2025, referred to as Traditional Current in the table. We use 30 evasive multiprocess ransomware samples~\cite{de_gaspari_evading_2022}, and three adaptive ransomware families designed specifically to evade \name totaling 20 samples (see Section~\ref{sec:adaptive_ransomware_generation} for details on adaptive families). Traditional Classic ransomware samples from~\cite{continella_shieldfs:_2016} were gathered in 2016, however, no details are available on when each sample was first seen in the wild. Evasive multiprocess ransomware are research prototypes and have no ransomware families. We use data from all proposed evasive configurations: process splitting, functional splitting, and mimicry~\cite{naked_sun_acns,de_gaspari_evading_2022}. Adaptive multiprocess ransomware are derived from the mimicry evasive multiprocess ransomware~\cite{de_gaspari_evading_2022} and are designed specifically to evade \name. We divide them into families based on the group of \name features that are manipulated to attempt evasion, as presented in Section~\ref{sec:adaptive_eval}.

Table~\ref{tab:dataset_benign} provides information on the benign dataset. The dataset of benign operations used in this study was provided by~\cite{continella_shieldfs:_2016}. The dataset was collected from 11 desktop computers over multiple weeks. It comprises computers dedicated to three different types of use: development (dev), office, and home. The dataset includes a wide variety of applications, including office suite applications (.doc, .ppt,  .xls), development tools such as Visual Studio, and compression utilities such as WinRar~\cite{continella_shieldfs:_2016}, providing a varied and challenging setting for behavioral modeling.

\begin{table}[t]
    \small
    \centering
    \begin{tabular}{ll|c|c}  \hline
     \multicolumn{2}{l|}{\textbf{Ransomware}} & \textbf{Year} & \textbf{No. Samples} \\\hline
     \multicolumn{2}{l|}{\textbf{Traditional Classic}} &  & 383 \\
        & Crypto Wall   & 2016  & 157   \\
        & Crowti        & 2016  & 125   \\
        & Crypto Defense& 2016  & 77    \\
        & CTB Locker    & 2016  & 14    \\
        & TeslaCrypt    & 2016  & 10    \\\hline
     \multicolumn{2}{l|}{\textbf{Traditional Current}} &  & 43 \\
        & Akira         & 2023-25  & 10   \\
        & LockBit       & 2023-25  & 10   \\
        & Conti         & 2023-25  & 10   \\
        & AvosLocker    & 2023-25  & 10   \\
        & WannaCry      & 2023-25  & 3    \\\hline
    \multicolumn{2}{l|}{\textbf{Evasive Multiprocess}} & / & 30 \\\hline
    \multicolumn{2}{l|}{\textbf{Adaptive}} & & 20 \\
        & Adaptive Entropy   & / & 5\\
        & Adaptive Op. Num.  & / & 9\\
        & Adaptive R/W Ratio & / & 6\\\hline
    \multicolumn{3}{l|}{\textbf{Total}}  &  476\\
    \hline
    \end{tabular}
    \caption{Ransomware families used in the evaluation. }
    \label{tab:dataset_ransomware}
\end{table}

\begin{table}[t]
    \centering
    \resizebox{\columnwidth}{!}{
    \begin{tabular}{c|c|c|c|c|c|c|c}  \hline
    \textbf{Machine} & \textbf{Win. Ver.} & \textbf{Type} & \textbf{Data (GB)} & \textbf{IRPs (Mln)} & \textbf{Proc. (Mln)} & \textbf{Apps} & \textbf{Duration (h)}\\\hline
    1   &   10  &   dev    & 3.4 & 230.8  & 16.6   & 317 & 34\\
    2   &   8.1 &   home   & 2.4 & 132.1  & 9.67   & 132 & 87  \\
    3   &   10  &   office & 0.9 & 54.2   & 5.56   & 225 & 17  \\
    4   &   7   &   home   & 4.7 & 279.9  & 18.70  & 255 & 122 \\
    5   &   7   &   home   & 2.2 & 138.1  & 5.04   & 141 & 47  \\
    6   &   10  &   dev    & 1.8 & 100.4  & 10.30  & 225 & 35  \\
    7   &   8.1 &   dev    & 0.8 & 49.0   & 3.28   & 166 & 8   \\
    8   &   8.1 &   home   & 0.8 & 43.9   & 6.33   & 148 & 32  \\
    9   &   8.1 &   home   & 7.7 & 501.8  & 24.20  & 314 & 215 \\
    10  &   7   &   home   & 0.9 & 57.6   & 2.63   & 151 & 18  \\
    11  &   7   &   office & 2.6 & 175.2  & 4.69   & 171 & 28  \\
    \hline
    \multicolumn{3}{c|}{\textbf{Total}}  & 28.2 & 1,763.0 & 107.00 & 2245 & 643\\
    \hline
    \end{tabular}
    }
    \caption{Benign dataset details.}
    \label{tab:dataset_benign}
\end{table}

\subsection{Ransomware Data Collection}
\label{sec:ransomware_collection}
We collected ransomware data on a Windows 11 virtual machine populated with $\sim33,000$ files distributed over various subdirectories within the \textit{User} Windows directory. We included over 80 different types of files, including different image formats (.jpg, .jpeg, .png, .gif), audio/video (.mp3, .mp4, .mov), compressed archives (.zip, .gz), and office files (.docx, .xlsx, .pptx). After each ransomware execution, we verified that the user files were successfully encrypted and discarded samples that did not correctly execute. The VM was reverted and reset after each execution. In total, we gathered 43 working samples from 5 different families, as illustrated in Table~\ref{tab:dataset_ransomware} (Traditional Current).

\begin{table}[t]
\caption{Detection rate of ShieldFS, RWGuard, and \name against traditional and multiprocess ransomware. Results taken from~\cite{continella_shieldfs:_2016,de_gaspari_evading_2022}.}
\centering
\begin{tabular}{c | c | c}
\toprule
Approach & Traditional $TPR$ & Multiprocess $TPR$ \\ 
\midrule
ShieldFS~\cite{continella_shieldfs:_2016}    &	100.00   &   0.00 \\
RWGuard~\cite{mehnaz_rwguard}     &	98.45   &   0.00 \\
\textbf{\name(ours)} &   99.25   &   99.98 \\
\end{tabular}
\label{tab:comparison}
\end{table}

\subsection{User Study Details}
We recruited five volunteer bachelor and master computer science students to carry out our user study. Each student was allowed to freely interact with two Windows systems deployed in a virtual machine: (1) standard system and (2) \name system. To ensure that key actions were evaluated, the users were asked to include in their workflow activities typically carried out by different types of users. Participants were asked to include the use of a web browser (standard user activity), edit documents on the local machine (office activity), perform software installation and removal operations (administration activity), and carry out file transfer operations. After interaction with each system, they were asked to rate the system usability on a scale from 1-5, considering in particular whether the system was responsive to interaction and whether activities were unusually slow.

\section{Comparison with SOTA}
We compare our approach against two state-of-the-art ransomware detection approaches: ShieldFS~\cite{continella_shieldfs:_2016} and RWGuard~\cite{mehnaz_rwguard}. As highlighted in recent publications~\cite{de_gaspari_evading_2022,zhou2023limits}, ShieldFS and RWGuard still represent reference benchmarks for behavioral-based ransomware detection. Table~\ref{tab:comparison} presents our comparison. As illustrated in the table, \name achieves performance comparable to state-of-the-art approaches against traditional ransomware. Furthermore, \name successfully detects evasive multiprocess ransomware, a feat that existing approaches fail to accomplish.

\section{Unseen Detection Performance}\label{sec:unseen_rw_evaluation}
Given the constant evolution of ransomware, an essential evaluation criterion for ransomware detectors is their ability to generalize to \textit{previously unseen} data. We assess \name's capability to generalize to unseen ransomware by randomly removing $10\%$ of the variants from our training data and evaluating detection performance on these withheld samples.
This evaluation provides a measure of the expected predictive and generalization capabilities of \name on future, unknown ransomware variants. Table~\ref{tab:unseen_performance} reports 
the results of our test. As highlighted in the table, \name's performance on unseen variants closely aligns with its performance on known variants, exhibiting only a minor performance degradation of ${\sim}0.2$ percentage points across all classifiers. These findings underscore the ability of \name to learn descriptive features that accurately capture the diverse range of behaviors displayed by ransomware.

\begin{table}[t]
\small
\caption{Performance analysis of \name against unseen ransomware families.}
\centering
\begin{tabular}{c | cccccc}
\toprule
 & Benign & Ransomware & \multicolumn{4}{c}{Overall}\\
\toprule
{Window} & $TNR$ & $TPR$ & Acc. & Prec. & Recall & F1 \\ 
\cmidrule(l){1-7}
0.25sec &	98.49 &  97.73 &	98.11 &	98.48 &	97.73 &	98.11 \\
0.5sec  &	98.45 &  98.44 &	98.45 &	98.45 &	98.44 &	98.45 \\
1sec    & 	98.41 &  98.62 &	98.52 &	98.41 &	98.62 &	98.52 \\
2sec    & 	98.65 &  98.79 &	98.72 &	98.66 &	98.79 &	98.72 \\
4sec    & 	99.09 &  98.92 &	99.00 &	99.08 &	98.92 &	99.00 \\

\midrule
stacked &   97.59 &  99.07 &	98.33 &	97.63 &	99.07 &	98.35 \\ 
\end{tabular}
\label{tab:unseen_performance}
\end{table}

\end{document}